# Interaction of Proteins with Polyelectrolytes: A Comparison between Theory and Experiment

*Xiao Xu[1], Stefano Angioletti-Uberti[2,3], Yan Lu[4,5], Joachim Dzubiella[4,6], Matthias Ballauff[4,7]*

[1]School of Chemical Engineering, Nanjing University of Science and Technology, 200 Xiao Ling Wei, Nanjing 210094, P. R. China

[2]Department of Materials, Imperial College London, London, UK

[3]International Research Centre for Soft Matter, Beijing University of Chemical Technology, 100099 Beijing, PR China,

[4]Soft Matter and Functional Materials, Helmholtz-Zentrum Berlin für Materialien und Energie GmbH, 14109 Berlin, Germany,

[5]Institute of Chemistry, University of Potsdam, 14467 Potsdam, Germany

[6]Physikalisches Institut, Albert-Ludwigs-Universität, 79104 Freiburg, Germany

[7]Institut für Physik, Humboldt-Universität zu Berlin, 12489 Berlin, Germany

AUTHOR EMAIL ADDRESS (matthias.Ballauff@helmholtz-berlin.de; joachim.dzubiella@physik.uni-freiburg.de )

ABSTRACT We discuss recent investigations of the interaction of polyelectrolytes with proteins. In particular, we review our recent studies on the interaction of simple proteins such as human serum albumin (HSA) or lysozyme with linear polyelectrolytes, charged dendrimers, charged networks, and polyelectrolyte brushes. In all cases discussed here we combined experimental work with molecular dynamics (MD) simulations and mean-field theories. In particular, isothermal titration calorimetry (ITC) has been employed to obtain the respective binding constants $K_b$ and the Gibbs free energy of binding. MD-simulations with explicit counterions but implicit water demonstrate that counterion release is the main driving force for the binding of proteins to strongly charged polyelectrolytes: Patches of positive charges located on the surface of the protein become a multivalent counterion of the polyelectrolyte thereby releasing a number of counterions condensed on the polyelectrolyte. The binding Gibbs free energy due to counterion release is predicted to scale with the logarithm of the salt concentration in the system which is verified both by simulations and experiment. In several cases, namely for the interaction of proteins with linear polyelectrolytes and highly charged hydrophilic dendrimers the binding constant could be calculated from simulations in very good approximation. This finding demonstrated that in these cases explicit hydration effects do not contribute to the Gibbs free energy of binding. The Gibbs free energy can also be used to predict the kinetics of protein uptake by microgels for a given system by applying dynamic density functional theory. The entire discussion demonstrates that the direct comparison of theory with experiments can lead to a full understanding of the interaction of proteins with charged polymers. Possible implications for applications as, e.g., drug design, are discussed.

KEYWORDS: protein, polyelectrolyte, calorimetry, simulation, ITC, HSA, lysozyme, dynamic density functional theory

1. INTRODUCTION

The interaction of proteins with polyelectrolytes is a long-standing and central subject in polymer and colloid science. Normally, proteins can carry both positive and negative charges on their surface and their overall charge depends on the pH in solution. Hence, electrostatic interactions play a major role when considering the binding of proteins to polymers or to charged surfaces in aqueous systems: First of all, proteins can interact with linear polyelectrolytes in aqueous solution to form complexes which may precipitate or generate a phase separation. These "complex coacervates" belong to the best-studied systems in general protein science and research in this field dates back to the twenties of the last century.[1–3] More recently, formation of complexes between proteins and polyelectrolytes was intensively studied by Dubin et al.[2,4–14], de Vries at al.[1,15–18] and by Kabanov et al.[19–22] Many of these investigations consider the interaction of polyelectrolytes with proteins carrying a net charge of opposite sign.[2] In these cases the strong interaction can be explained by the electrostatic monopole-monopole attraction in these systems. Thus, a cationic polyelectrolyte interacts with a protein at a pH above its isoelectric point and vice versa. Careful work, notably done in the group of Dubin et al.[4,5,7,8,10,23] has revealed, however, that interaction can occur at the "wrong side" of the isoelectric point, that is, a polyelectrolyte binds to a protein having the same charge. Here a force must be operative that overcomes the strong monopole-monopole repulsion. This finding hence demonstrates that a quantitative understanding of the interaction between proteins and polyelectrolytes requires an in-depth discussion of all possible driving forces which will be done further below.

Linear polyelectrolyte chains can be assembled to form charged polymeric gels. The interaction of proteins with these charged macroscopic gels or microgels presents an equally well-studied subject[20,21,24–34] which finds practical application, e.g., in the chromatographic purification of proteins.[35–38] DNA and RNA form strong complexes with branched cationic polyelectrolytes and these complexes are used for gene transfection and gene therapy.[32,33,39] Charged dendrimers[40,41] have been tested for all kinds of biomedical applications[42,43] and the interaction with proteins is of central importance in the field.[44] The obvious importance of this problem has led to an enormous literature that is hard to overlook. Last not least, linear polyelectrolytes may be densely grafted to planar and curved surfaces.[45,46] The polyelectrolyte chains and brushes thus obtained interact strongly with proteins in aqueous solution and form a variety of protein-polyelectrolyte assemblies.[3,47–50] Such systems are of general importance in nanotechnology since polymer chains are often used to prevent the adsorption of proteins onto nanoparticles.[51] Proteins that adsorb to nanoparticles, e.g., in the blood stream may denaturate and thus trigger an immune response of the body.[52–61] Hence, nanotoxicology[62,63] that deals with these adverse effects of nanoparticles must consider the interaction of proteins with such polymeric layers. Knowledge on this interaction will ultimately contribute to a better understanding of nanoparticles in complex biological media and their ultimate fate within the body.[64,65]

This short overview demonstrates the extraordinary breadth of the field in which the interaction of polyelectrolytes and proteins will matter. Evidently, an advanced understanding of the driving forces leading to formation of protein/polyelectrolyte complexes would be of enormous value inasmuch it would allow us to predict the association strength of such objects and a quantitative calculation of the binding constant. Obviously, a quantitative treatment of this interaction must start on the molecular level where a single charged unit is adsorbed onto the surface of a protein. Theory must hence advance with sufficient resolution and consider all pertinent length scales from molecular to mesoscopic distances. At the same time, understanding the interaction of small charged molecules with proteins is central for modern drug design since many drugs represent charged entities. Moreover, small charged molecules as, e.g., phenylacetic acid or indoxylsulfate present toxins that adhere to blood proteins such as human serum albumin.[66–69] Full removal of these toxins by dialysis presents a central and urgent problem of modern nephrology.[68]

Precise thermodynamic information on the binding process is the experimental prerequisite for a detailed understanding of the interaction of proteins with polyelectrolytes. In approximately the last two decades, isothermal titration calorimetry (ITC) has brought tremendous progress in this field.[70–77] ITC measures the heat when, e.g., a solution of a polyelectrolyte is titrated into a solution of a protein. If the caloric effect is high enough, this method yields the overall heat of binding $\Delta H_{ITC}$ and the binding constant $K_b$ that is directly coupled to the change of Gibbs free energy of binding $\Delta G_b$. This quantity may be used in turn to derive the entropy of binding $\Delta S_b$ by measuring its dependence on temperature $T$.[78–80] Thus, ITC allows us to get the full thermodynamic information on the binding process, in particular on the interaction of biological polyelectrolytes with proteins.[78–82] Hence, by now there is a broad set of thermodynamic data in literature obtained from a wide variety of systems. General trends may now be discussed as, e.g., the role of enthalpy in the binding of drugs to proteins.[76] On the other hand, research by ITC has led to a number of controversies and open questions.[76] Thus, very often there is a nearly full cancellation of enthalpy and entropy in binding processes conducted in aqueous medium. This "enthalpy-entropy cancellation" (EEC) presents a general phenomenon and has led to an intense discussion in literature (see, e.g., the discussion in ref.[76]).

The spatial structure of complexes of proteins with polyelectrolytes is another central problem in the field. Small-angle scattering as small-angle x-ray scattering (SAXS)[83–86] and small-angle neutron scattering[87] (SANS) have become important tools for their elucidation. Progress in synchrotron instrumentation allows us now to follow the build-up of the complexes in a time-resolved manner.[85] Static and dynamic light scattering have been applied to the analysis of complexes in solution. Fluorescence microscopy[61,88] and fluorescence spectroscopy[89] have been applied quite often, too, and it seems fair to state that we have acquired a rather detailed understanding of the spatial structure of the complexes in solution. Together with ITC-data these methods have led to a better understanding of a number of complexes. However, there is no general conclusion on the fundamental driving forces of complex formation nor a set of accepted theoretical models that allow us to truly predict the Gibbs free energy of binding for a given system.

In the last decade our groups have worked intensively on a quantitative understanding of the interaction of proteins with charged polymeric systems.[26,27,48,51,69,90–94] These investigations include i) the interaction of small charged molecules with human serum albumin[69,87] (HSA), ii) the interaction of short linear polyelectrolytes with HSA,[91] iii) the interaction of charged dendritic polymers with lysozyme and HSA[94,95], iii) the interaction of various proteins with charged microgels,[24–28,92,96] and iv) spherical polyelectrolyte brushes interacting with various proteins.[46–48,90,97–100] The main goal of our work is the elucidation of the various driving forces leading to complexation between proteins and polyelectrolytes. Central to this question is the role of the electrostatic interaction as opposed to other forces as, e.g., the hydrophobic interaction. Thus, we have conducted a series of experiments as a function of the ionic strength and temperature in solution. The central point in this research is the direct comparison of all experiments with theory and simulations.[91,94] Realistic models for polyelectrolytes and for proteins allows us to investigate complexation in a fully quantitative fashion. All simulation models treat the counterions in an explicit fashion and thus lead to a quantitative treatment of the entropic contribution to binding by the release of counterions:[69,90,94,101,102] Proteins may carry patches of positive charge on their surface even above the isoelectric point where the overall charge is negative. These patches may act as multivalent counterions to the polyelectrolytes thereby releasing the condensed counterions of the highly charged chains. This *counterion release force*[102] can assume an appreciable magnitude depending on the charge density of the polyelectrolyte and the salt concentration in bulk. Also, it is still operative under physiological conditions, that is, at high ionic strength. A clear proof for the importance of the counterion release force could be given for the interaction of linear polyelectrolyte with proteins[91,94,98,103] and for the binding of proteins onto spherical polyelectrolyte brushes.[93,98]

Our recent work on the interaction of HSA with short chains of poly(acrylic acid)[91] has demonstrated that simulations can predict $\Delta G_b$ quantitatively with surprising accuracy. The same finding was made when considering the interaction of a highly charged dendritic polymer with HSA or with lysozyme.[94] These surprising results opens the door for a fascinating development towards a design of polymeric systems that interact with proteins in a prescribed manner. Hence, future investigations must aim at this fully quantitative and predictive modeling of the interaction of proteins with polyelectrolytes.

In this review we present recent work dealing with the interaction of proteins with polyelectrolyte systems with special emphasis on the quantitative comparison of theory and experiment. Thus, we first discuss theoretical models which are compared to experimental data in turn. In particular, we discuss a number of systems where the protein and the polyelectrolyte have the same charge, that is, where the interaction takes place on the "wrong side" of the isoelectric point. The review is organized as follows: First, the interaction of a protein with a linear polyelectrolyte is considered.[87,91,101] In addition to this, we discuss recent work on the interaction of a highly charged dendrimer[41] with lysozyme.[94] In a next step we review the interaction of proteins with weakly charged networks.[28,92] Here weakly charged means that no counterion condensation occurs on the chains constituting the network. Then we shall present a survey on recent work done on spherical polyelectrolyte brushes which are characterized by a strong confinement of the counterions.[93,104] Up to this point the discussion is restricted to full equilibrium and an entirely thermodynamic treatment. Recent theoretical work, however, has suggested that the kinetics of the binding of proteins to gels can be understood in terms of a dynamic density functional theory.[92,105] Within this theory, the thermodynamic driving force in this model can be directly taken from the previous equilibrium considerations. A brief conclusion section will wrap up the entire discussion.

2. PROTEINS INTERACTING WITH LINEAR AND DENDRITIC POLYELECTROLYTES
   *2.1 Linear polyelectrolytes: Theory*

As mentioned above, the interaction of linear polyelectrolytes with proteins in aqueous solution belongs to the best-studied field in colloid science.[1,2,13,77] In the following we only consider the interaction of a negatively charged polyelectrolyte chain in aqueous solution with a protein above its isoelectric point. Both objects hence carry a net negative charge and should repel each other. Work by Dubin and coworkers[8,13] and by us[91], however, clearly demonstrated an attractive interaction of the protein with the a highly charged polyelectrolyte if the ionic strength is low. The analysis of this effect by isothermal titration calorimetry (ITC) clearly revealed the entropic origin of this interaction.[91,98,103] This can be argued directly from the fact that the enthalpy measured by ITC is strongly positive. This entropic attraction is mainly due to the counterion release force which is directly related to the uneven charge distribution on the surface of proteins.[102] As an example, Figure 1a displays the surface distribution of charges of beta-lactoglobulin. The overall charge is negative since the pH in solution was set above the isoelectric point of the protein. However, the blue spots mark the patches of positive charge which can interact with the negatively charged polyelectrolyte. Their binding then releases $\Delta N_-$ negative counterions of the positive patch as well as $\Delta N_+$ positive counterions of the polyelectrolyte. The release of counterions thus effectively increases the entropy of the system and leads to a concomitantly negative contribution to the Gibbs free energy of binding. More specifically, the change $\Delta G_{cr}$ can be estimated by[90,98]

$$\frac{\Delta G_{cr}}{kT} = \Delta N_- ln\left[\frac{c_s}{c_{patch}}\right] + \Delta N_+ ln\left[\frac{c_s}{c_{PE}}\right] \quad (1)$$

where $c_s$ is the salt concentration in the solution, $c_{patch}$ the concentration of the negative counterions accumulated on the positive patch of the protein, $c_{PE}$ is the surface concentration of the condensed counterions of the linear polyelectrolyte and $k$ and $T$ have their usual meanings. Evidently, $\Delta G_{cr}$ may assume large negative values depending on the ratio of the concentrations $c_{patch}$ and $c_{PE}$ to the salt concentration $c_s$ in the bulk solution.

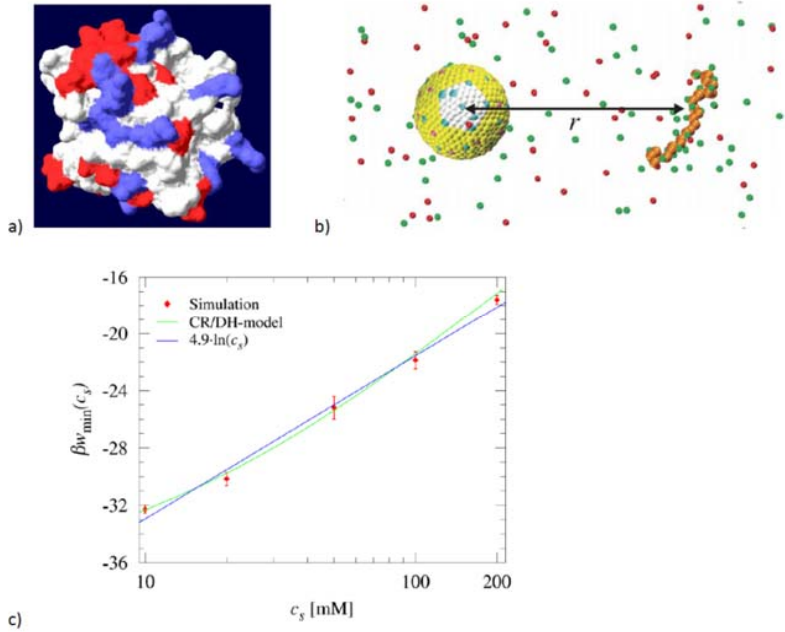

**Figure 1**. Interaction of linear polyelectrolytes with proteins. Fig. 1a: Charges on the surface of ß-lactoglobulin (BLG). Red: negative (acidic) residues; blue: positive (basic) residues; white: neutral residues. The overall charge of the protein is negative since the pH is above the isoelectric point pI. Fig. 1b: Simulation of the interaction of a linear polyelectrolyte with a protein modeled by the charged patchy protein model (CPPM).[101] The protein is modeled by a sphere with a single patch of positive charge (green dots on white spot) bearing 8 positive charge. The overall charge is -8. The linear polyelectrolyte models a short chain of poly(acrylic acid) with 25 repeating units. Each unit carries one charge. Note that the counter- and the coions are modeled explicitly.[101] Fig. 1c: Free energy of binding $w_{min}$ of a polyelectrolyte to a patchy particle as the function of salt concentration $c_s$. The points are derived from the CPPM-simulations whereas the blue solid line denotes the fit according to eq.(1). The green line displays the result of the analytical model derived in ref.[101]

It is often assumed that the electrostatic interaction is not operative under physiological condition where $c_s$ is of the order of 150 mM which is followed by a very small Debye-length defined by $\kappa^{-1} = (8\pi l_B)^{-1/2}$ with the Bjerrum-length $l_B = e^2/(4\pi\varepsilon_0\varepsilon_r k_B T)$. Here $e$ denotes the unit charge, $\varepsilon_0$ is the permittivity of the vacuum and $\varepsilon_r$ is the dielectric constant. However, depending on the charge density of the protein and the polyelectrolyte, the surface concentration may assume values of the order of several moles per liter which leads to a still appreciable $\Delta G_{cr}$ even under physiological conditions. On the other hand, the effect will vanish if $c_s$ is of the order of $c_{PE}$ or $c_{patch}$. Eq.(1) also suggests that the free energy of binding should scale with $ln(c_s)$ if the counterion release force is operative.[102] The derivative of $\Delta G_{cr}$ with regard to $ln(c_s)$ hence yields directly the number of release counterions. This feature has been repeatedly found in experimental studies and will be further discussed below.

The number of condensed ions on a PE chain can be well estimated by the Onsager-Manning-Oosawa theory.[106–109] The linear polyelectrolyte is characterized by the Manning parameter $\xi$

$$\xi = z l_B / l \quad (2)$$

where $l$ is the distance between the charges along the polyelectrolyte chains whereas $z$ is the valency of the counterions. Theory predicts that counterion condensation starts for $\xi > 1$. For vanishing salt concentration the fraction $x$ of condensed counterions follows as $x = 1-1/\xi$, that is, a fraction x of the counterions accumulates in the immediate vicinity of the linear polyelectrolyte chains. The concentration of the condensed counterions that exhibits only a weak dependence on $c_s$ can be estimated for typical polyelectrolytes as, e.g., poly(acrylic acid) from simulations[91,110] to be ca. 1.5 M. A recent estimate[90] for $c_{patch}$ for a planar spot with a typical charge density led to approximately 0.3 M which will be even reduced when considering the curved surface of proteins.[101] Hence, it suffices in good approximation to consider only the term related to $c_{PE}$ in eq.(1).[101]

The above considerations can be put into more quantitative terms by using the charged patchy particle model (CPPM).[111,101] Figure 1b displays the main feature of this model:[101] The proteins are rendered by spheres of 2 nm diameter. The negative charges marked in red are distributed over the surface of the sphere at random while the positive surface charges marked in green are concentrated in a patch marked in white. The charge density of such a patch is of the order of $1 - 2$ e/nm$^2$ which is typical for globular proteins of this size. The patchy distribution also leads to a marked dipole moment of the order of $10^2$-$10^3$ Debye which is typical for proteins. The net charge was chosen to be -8 in order to mimic proteins above the isoelectric point. The polyelectrolyte is model in a coarse-grained fashion by 25 beads each of which carrying one charge. The solvent water is modelled by a background continuum with $\varepsilon_r$ =78 whereas all counter- and coions are treated explicitly. The overall interaction of the protein with the polyelectrolyte is obtained by a *steered Langevin simulation*.[111,101] Briefly, in this simulation the center of gravity of the polyelectrolyte is slowly moved towards to the protein along a prescribed direction. At each point the forces are sampled and added up to give the *potential of mean force* (PMF) as the function of the distance *r*. The simulation carries along all pertinent interactions as the van der Waals forces, as well as dipolar interactions and Coulombic interaction between all beads. Moreover, the explicit treatment of all ions leads to the number of condensed counterions that are located in a vicinity of 0.4 nm to the surface of the protein or the polyelectrolyte. All other counterions are considered to be free. [101]

The main result of this model is the clear proof of an attractive interaction due to the release of counterions, mainly the ones bound to the polyelectrolyte. The polyelectrolyte chain is seen to be directly attached to the patch and depending on the size of the patch the chain may be even adsorbed totally. Figure 1b displays a typical configuration. The PMF exhibits a distinct minimum of the order of -20 $k_BT$ and the dependence on the salt concentration scales essentially as predicted by eq.(1) (see Figure 1c). Both the counterion release and the screening effects can be treated in an analytical model which is capable of fully describing the simulations (green line in Fig. 1c).

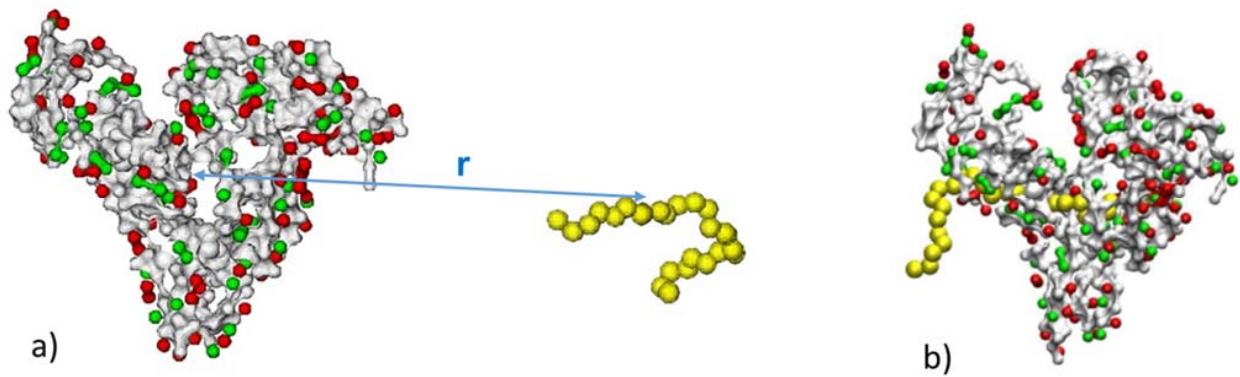

**Figure 2** Computer simulations of the binding of poly(acrylic acid) (PAA) to HSA. a) Steered Langevin Dynamics Simulation of the binding of PAA to HSA. The center of mass of the PAA-molecule is kept at a given distance to the center of mass of HSA and moved towards the protein by a constant pulling speed. The forces acting on the PAA-molecule are averaged and integrated up to get the potential of a mean force (PMF) which in turn leads to $G(r)$. The simulated Gibbs free energy of binding $G^{sim} = G(r_{min}) - G(\infty)$ where $G(r_{min})$ denotes the minimum value of $G(r)$ at the bound state. See ref.[91] for further details. b) Simulation snapshot of the complex of PAA and HSA. One PAA-molecule is bound to the Sudlow II site of HSA.[91]

A more detailed and specific analysis has been done on the interaction of human serum albumin (HSA) with poly(acrylic acid) (PAA) by a coarse-grained computer model.[91] Here the protein has been modeled in terms of a *Go-model* provided by the SMOG webtool for biomolecular simulations.[112,113] The amino acids are modeled by single beads. The short flexible PAA-chain is again modeled in a coarse-grained fashion as in the above simulations and its interaction with HSA is determined in a steered Langevin simulation.[114] Water is treated implicitly as a continuum with a dielectric constant depending on temperature.

Figure 2 shows representative snapshots of these simulations. First of all, the simulations clearly indicate that only one PAA-chain is interacting with a HSA molecule. Moreover, the simulation suggests the exact location of the binding of the polyelectrolyte, namely the Sudlow II site. The PAA-chain is seen to slide along this side much in a way of threading through an orifice. Evidently, the free ends of the chains on both sides are free to explore all possible configurations which increases the entropy of this state. Thus, this configurational degree of freedom at the Sudlow II site certainly favors this site over other positive patches on the surface of HSA. This fact is one of the main results of the simulations[91] and demonstrates that coarse-grained models can give highly detailed information at the pertinent length scale.

As delineated above, the steered Langevin simulations allow us to obtain the free energy profiles *G(r)* by integrating the average forces along the distance coordinate between the centers of gravity of HSA and PAA. The minimum of *G(r)* at this position gives directly the free energy of binding which can be obtained at different salt concentrations and temperatures. The simulations reveal that counterion release is the main driving force and 2.5 ions are released upon binding. Deeper analysis of the simulation data show that ca. 2 of these ions stem from the polyelectrolyte. Thus, with an approximate concentration $c_{PE}$ of 1.5 M eq.(1) gives a free energy of the order of $3 - 4\ k_BT$ per ion.

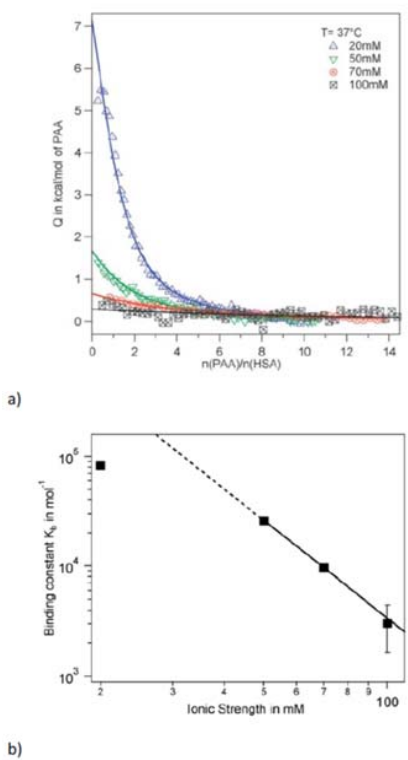

**Figure 3** Analysis of the binding of poly(acrylic acid) to HSA in aqueous solution at 37°C. a) Binding isotherms measured by ITC for different salt concentrations. The heat of dilution has been subtracted from the ITC-signals. The solid lines show the respective fits of the data by the single set of independent binding sites (SSIS) model. b) The dependence of the Gibbs free energy of binding $\Delta G_b = -kT \ln K_b$ on salt concentration $c_s$. The line is the fit according to equation (1). Taken from ref.[91]

*2.2 Experiment*

ITC was used to determine the Gibbs free energy of binding experimentally.[91] Figure 3a gives as an example the measured heat corrected for the heat of dilution. The data have been taken at 37°C and the data show clearly that the process of binding is endothermic, that is, entropy is the driving forces for complex formation as has already been found in earlier studies.[98,103] The solid lines describe the binding isotherm modeled as a simple chemical equilibrium in terms of an equilibrium constant $K_b$. Hence, $\Delta G_b = -kT \ln K_b$ leads directly to the Gibbs free energy of binding $\Delta G_b$ that can be compared to the values obtained from simulations.[91]

The comparison of simulated and measured free enthalpies of bindings requires special consideration.[91,94] The $\Delta G_b^{exp}$ obtained from ITC experimentally refers to the standard volume $V_0$ of one liter per mole. In the PMF simulations, the binding volume $V_b$ is given by the volume accessible to the center of mass of PAA in the bound state. Hence, a term $\Delta G_b^{corr} = -k_B T \ln(V_B/V_0)$ must be added to $\Delta G^{sim}$ in order to obtain the *standard* Gibbs free energy of binding that can be compared to experimental values (cf. also the discussion of this problem in ref.[115]).

Excellent agreement between the measured and the simulated $\Delta G_b$ was found. Moreover, it was found that counterion release is indeed the driving force for binding. Figure 3b displays a fit of $\Delta G_b$ according to eq.(1). As already argued above, only the second term of eq.(1) needs to be taken into account since the counterion condensation on the charged patches of the proteins is negligible. Figure 3b demonstrates that the linear relation between $\Delta G_b$ and $\ln(c_s)$ is found indeed except for the lowest salt concentration. This deviation from linearity has been found before and can be explained by the repulsive monopole-monopole interaction between the negatively charged protein and the anionic polyelectrolyte. Thus, if the salt concentration becomes very low, this repulsion is not screened anymore and the magnitude of $\Delta G_b$ decreases when compared to the prediction of eq.(1).

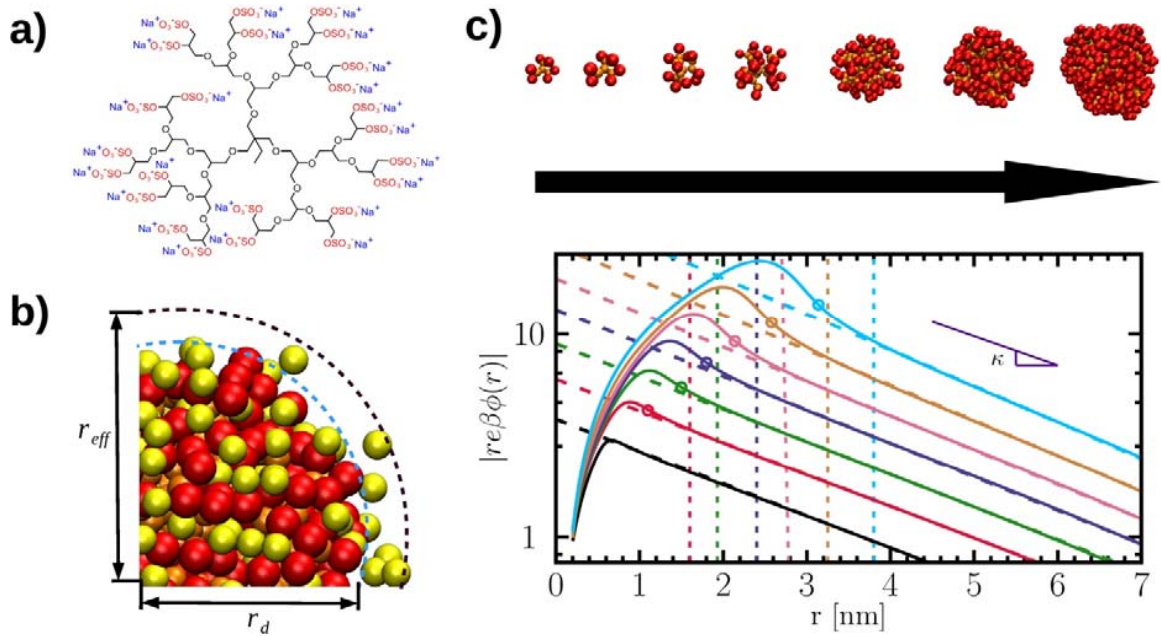

**Figure 4** Spatial structure and counterion condensation for highly charged sulfated polyglycerol dendrimers (dPGS).[41] 4a) Chemical structure of a PGS-dendrimer of second generation. 4b) The illustration of the bare and effective radius of the dPGS concerning the snapshot of the G4-dPGS molecule in the coarse-grained fashion. The orange and red beads denote the branching and terminal repeating segments of dPGS, whereas the yellow spheres represent the condensed counter ions. 4c) Simulation snapshots and scaled electrostatic potential of a PGS-dendrimers as the function of distance for generation 0 (black) to 6$^{th}$ generation (turquoise). Lines with the slope κ reflect the expected Debye-Hückel decay at long distances.

## 2.3 Dendritic polyelectrolytes and proteins

Dendrimers are branched polymers with a regular structure. Starting from a focus point 3 – 4 branches are emanating that become the focal point for the next generation.[40] Hyperbranched polymers have a similar but less well-ordered structure. Charges can be appended to each repeating unit or just to the terminal units. Dendritic polymers have been intensively discussed for various medical applications as e.g. drug delivery.[42,43] This application necessitates precise information on the interaction of charged dendrimers with proteins in solution. Here a problem of central importance is the possible degradation of the secondary and tertiary structure of the protein in the complex which virtually excludes the medical use of the respective dendritic polymer. Several studies of this problem have indicated that dendrimers can lower the stability of proteins considerably. Thus, CD-spectroscopy of, e.g., the complex HSA with PAMAM dendrimers demonstrates that the melting of the secondary structure of HSA occurs in the complex at considerably lower temperature than measured for the free protein.[60,116,117]

Another central problem in the field is the use of dendrimers as drug. Some time ago it has been demonstrated that dendritic polyglycerol sulfate (dPGS) is an anti-inflammatory drug.[118–120] Figure 4a shows the chemical structure of a dPGS of second generation. Sulfonate groups are appended at the ends of each branch rendering the entire molecule a highly charged polyelectrolyte. This anti-inflammatory potential of dPGS was traced back to the blocking of selectins during the immune response, that is, it could be explained by the selective interaction of dPGS with a given protein.[119,121] Thus, dPGS forms a complex with L- and P-selectin but not with E-selectin.[119] Furthermore, early studies could show clearly that this specific interaction is directly related to the high charge of dPGS, uncharged poly(glycerol)s behave in a total different way.[122]

Recently, we have studied the interaction of human serum albumine (HSA) and lysozyme with dendritic polyglycerol sulfate (dPGS).[94,95] These proteins serve as well-defined and stable model systems for analyzing the details of the interaction with dPGS. We combined thermodynamic experiments using ITC with coarse-grained molecular dynamics simulations. All studies aimed at a quantitative assessment of the electrostatic part of the free energy of binding as opposed to factors related to hydration. Thus, the binding constant was determined experimentally at different temperature and salt concentrations and compared to MD-simulations. In a second step these results could be used for a quantitative modeling of the interaction of dPGS with selectins. In particular, the dependence of the binding constant on temperature has been used to derive the enthalpy $\Delta H_b$ of binding and the respective entropy of binding $\Delta S_b$.[95] Both quantities can then be related to the Gibbs free energy of binding $\Delta G_b$. Here we found a marked enthalpy-entropy cancellation (see ref.[76]), that is, a major part of $\Delta H_b$ is cancelled by $\Delta S_b$ while both quantities will even change sign within a small temperature range from 10 – 40°C. This strong EEC seen in many biological systems[76,80–82,123] seems to be a very general feature there. The system dPGS/HSA studied by us is a model system that allows us to study the EEC with sufficient precision.[95] The data obtained from this model systems hence allow us to discuss all thermodynamic functions and their relation on a secure basis and relate them to simulations.

***Theory***:

Highly charged dendritic structures have been the subject of many investigations since the nineties.[104] Thus, there is detailed knowledge about the well-studied PAMAM-dendrimers by simulations by now.[124–127] In our work[41] we employed MD-simulations using a coarse-grained model of dPGS and analyzed in particular the surface of dPGS, its location and the effective surface potential. This knowledge is central for a meaningful comparison with experimental data referring to measurements of the zeta-potential. With proper definition of the surface and the effective potential, the local concentration of condensed counterions can be given and used in calculations using equation (1). As shown in ref.[41] a rather well-defined core of dPGS with radius $r_d$ could be obtained from the simulations for generation 2 -5 which assume a more or less perfect spherical shape, in particular at higher generations. The outer surface at distance $r_{eff}$ in Figure 4c was obtained by mapping the calculated electrostatic potential onto the Debye-Hückel form. As a result of this modeling, the zeta-potential of dPGS could be determined by simulation for all generations and compared with experimental data and analytical theories. Good agreement was found indicating that the definitions of $r_d$ and $r_{eff}$ are giving a realistic picture of the spatial distribution of the counterions. In consequence, a simple but reliable model of the dPGS in solution could be set up shown in Figure 4b. The condensed counterions are located between $r_d$ and $r_{eff}$ and an effective surface concentration $c_{ci}$ of these ions can be calculated.[94] Thus, for a dPGS of second generation we obtained 0.96 M whereas a concentration of 2.43 M results for dPGS of fifths generation (see Table S3 of ref.[94]). These concentrations are much higher than e.g. the physiological salt concentration of 0.15 M and a release of counterions from the shell of the dPGS will lead to a marked gain of Gibbs free energy according to eq.(1). The strong localization of the counterions in this shell leads also to an effective charge per surface that increases only slowly with increasing generation. Concomitantly, the zeta-potential levels off at higher number of generation in good agreement with experimental data from the zeta-potential.

In a next step the binding of lysozyme to this well-characterized dendrimer was analyzed, again using the steered Langevin simulations already discussed above.[94] Figure 5 displays the main results of these simulations. Depending on the number of generations 3 to 14 lysozyme molecules are bound to dPGS and the upper part of Figure 5 displays typical snapshots of the first bound molecule. The middle of the panel Figure 5 shows the potential of a mean force for the first bound lysozyme, simulated for different generations. A distinct minimum indicates the strong driving force for complex formation and the lower panel displays the number of released counterions when one lysozyme is approaching a dendrimer. Here the yellow points mark the number of released counterions upon binding. The simulations again indicate very clearly that counterion release is the main driving force for complex formation. The origin is located in a patch of positive charge shown in the inset of the middle panel. The interaction of this patch with the highly charged surface is sufficient to release 3 -5 condensed ions from the dPGS and create a marked contribution to $\Delta G_{cr}$ according to eq.(1).

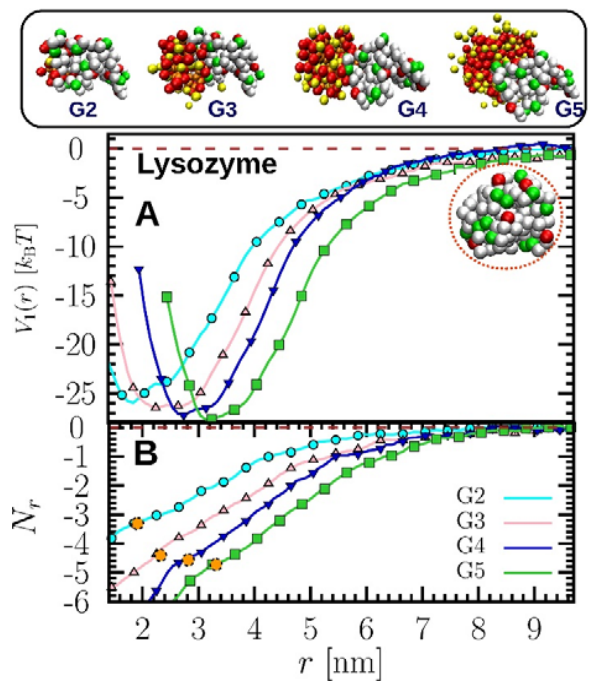

**Figure 5** Simulation of the binding of lysozyme to a highly charged dendrimer.[94] The upper panel shows snapshots of the first bound lysozyme to PGS-dendrimers from the second (dPGS-G2) to 5$^{th}$ generation (dPGS-G5). Panel A displays the potential of a mean force termed $V_1(r)$ for the first lysozyme molecule approaching a dPGS of different generations as indicated by the color code. Panel B shows the number of released counterions when the first lysozyme molecule is approaching the dendrimer. Yellow circles refer to the bound state. Taken from ref.[94].

These simulations allow us to analyze the complex formation of one dendrimer with several lysozyme molecules. Evidently, there must be a steric and electrostatic repulsion for each additional lysozyme molecule attaching to the surface of the dendrimer. This effect constitutes hence a negative cooperativity for the process of binding which in turn means that the Gibbs free energy of binding is not a constant but becomes an explicit function of the number of bound molecules. Many current models of modeling experimental data, on the other hand, assume a Langmuir-type model which proceeds from the assumption of a constant binding energy. The consequences for the evaluation of the ITC-data will be discussed in the next section.

*Experiment*
Figure 6 displays a survey of the binding studies carried out by ITC. Panel 6a displays the binding isotherms obtained from ITC. The fits have been done by a Langmuir model and the number of bound lysozymes can be obtained as a fit parameter. The binding constant $K_b$ can be converted into a Gibbs free energy of binding through $\Delta G_b = -k_B T \ln K_b$. The dependence on salt concentration is shown in Figure 6b referring to the complex formation of lysozyme with the dendrimer of second generation. The weakening of the binding can be directly seen from the inset displaying $\ln K_b$ as the function of salt concentration.

Good linearity is seen except for the smallest salt concentration. As already discussed in the case of HSA interacting with poly(acrylic acid), this deviation stems from a marked contribution of the screened electrostatic forces that are strongly operative under small salt concentration. The slope of the straight line leads to $N_{cr}$ =3.1 ± 0.1 in excellent agreement with the simulations.

As discussed above, the negative cooperativity can be seen directly in curves of the potential of a mean force $V_r(r)$ as the function of the center-of-mass distance r shown in panel Figure 5C for a dendrimer of 5$^{th}$ generation. The insets display snapshots of the growing corona of proteins. The deep minimum of the first bound lysozyme amounts to more than -25 $k_BT$. However, the minimum for the subsequent 14 lysozymes becomes smaller with each newly added molecule because of steric and electrostatic repulsion. Thus, the minimum for the 15$^{th}$ molecule is so small that no binding takes place anymore. A detailed analysis of the simulation data indicates that the coordination number of lysozyme on the dendrimer of 5$^{th}$ generation is 13 whereas the ITC experiments lead to 12 bound lysozymes. Considering the simplifications in the coarse-grained simulations this may be regarded as very good agreement.

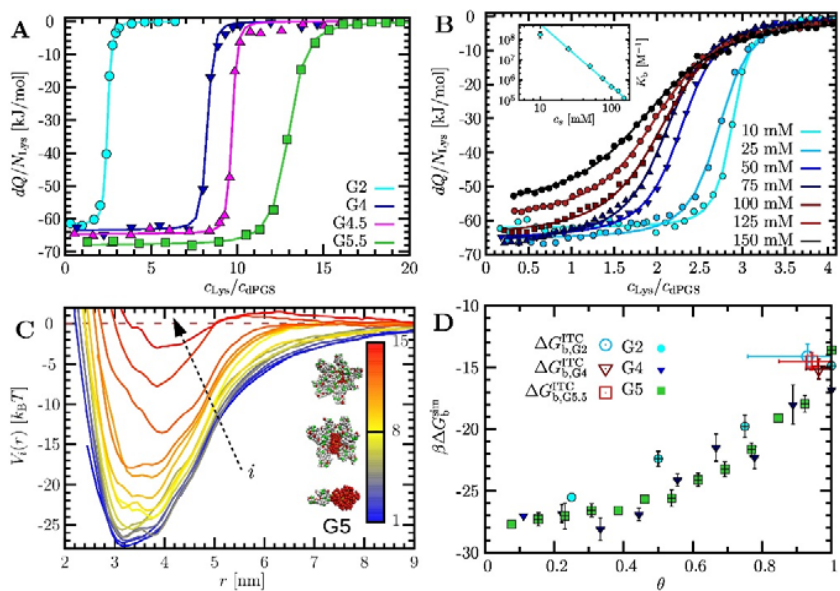

**Figure 6** Analysis of the binding of lysozyme to the highly charged dendritic polymer dPGS.[94] **Panel A** displays the ITC-isotherms for the binding of lysozyme to dPGS differing in the number of generations (see color code in Fig. 6A). Note that the dendritic polymers derive from a hyperbranched precursor and the number of generations indicated in the graph have been derived from the average molecular weights of the precursors (see ref.[41] for further details.) **B** Complexation of a dPGS-G2 (second generation) with lysozyme at different salt concentrations $c_s$. The solid lines display the fits by the Langmuir model. The inset shows the binding constant $K_b$ as the function of $\ln(c_s)$ as suggested by eq. (1). **C** The potential of mean force as the function of the center-of-mass distance $r$ between a dPGS-G5 and lysozyme for the successive binding of i = 1- 15 proteins simulated for a salt concentration of 10 mM. The inset shows snapshots of of the complexes resulting for i = 1, 8, and 13 bound proteins. **D** The simulation Gibbs free energy of binding in units of $k_BT$ (symbols) plotted against the degree of coverage $\theta = i/N_{sim}$ for dPGS-G2, G4, and G5, respectively. These data have been taken from the global minima of the PMF shown in

Figure 6C. The large open symbols display the respective simulation-referenced Gibbs free energy $\Delta G_b^{ITC}(i^*)$ (see eq.(3) and text) for comparison.

From these results the question arises how one could compare the simulated free energy with the data deriving from ITC-experiments. In case of a one-to-one-binding as discussed above for the complex formation of HSA with poly(acrylic acid), this comparison only must take into account the different volumes of binding. In case of multiple binding, however, the Gibbs free energy of binding depends on the number of already bound lysozymes as shown in Figure 6C. Figure 6D displays these minimum values $\Delta G_b^{sim}$ that can be read off directly from the minima of $V_i(r)$ in Figure 6C. Now the independent variable is the coverage $\theta$ that can be taken from the fits of the Langmuir model in Figure 6A. As discussed recently by us, the comparison of $\Delta G_b$ with $\Delta G_b^{sim}$ can be done by defining a *simulation-referenced Gibbs free energy*.[94] The binding constant follows from the slope of the ITC-isotherm directly at the inflection point. This point corresponds to a coordination number $i^*$ that is slightly below the maximum number $N$ of binding sites derived from the Langmuir fits. Hence, the coverage at the inflection point $\theta^* = i^*/N$ is smaller than unity. For a dendrimer of 5th generation, e.g., we find $\theta^* = 0.94$. The Gibbs free energy $\Delta G_b$ corresponds to this coverage $\theta^*$ and we may do the comparison through[94]

$$\frac{\Delta G_b^{ITC}(i^*)}{k_B T} = \frac{\Delta G_b}{k_B T} - \ln\left(1 - \frac{i^*}{N}\right) - \ln(\frac{NV_0}{V_B}) \qquad (3)$$

Here $\Delta G_b^{ITC}(i^*)$ is the simulation-referenced Gibbs free energy that can directly be compared with $\Delta G_b^{sim}$ whereas $\Delta G_b$ is the Gibbs free energy calculated from the binding constant $K_b$. The second term on the right-hand side is related to the entropic penalty for binding in the Langmuir model and the third term is the correction for the binding volume $V_b$ deriving from the PMF simulations as compared to the standard volume $V_0$ (1 l/mol) in the Langmuir approach.[115] Figure 6D displays the comparison of $\Delta G_b^{sim}$ and the simulation-referenced Gibbs free energy $\Delta G_b^{ITC}(i^*)$ for dPGS of second and of fifth generation. There is full agreement of simulations and experiment within the prescribed margins of error.

These studies have been continued by a comprehensive investigation of the binding of HSA to dPGS-G2, again combining analysis by ITC with MD-simulations for temperatures ranging from 278 to 313K.[95] First of all, the analysis by ITC demonstrated that HSA and dPGS-G2 form a 1:1 complex at all temperatures under consideration. This fact renders the subsequent discussion simpler. The binding can now be described in terms of a simple chemical equilibrium described by a single constant $K_b$. An analysis of the structure of HSA by CD-spectroscopy over the given temperature range furthermore corroborated that the heat signal measured by ITC is solely due to binding, no changes of the secondary or tertiary structure can be detected. The experimental Gibbs free energy $\Delta G_{b,ITC}$ derived therefrom can be compared directly to the simulated value $\Delta G_{b,sim}$. Practically full agreement is found for 283 and 298K whereas the MD-simulations slightly overestimate $\Delta G_b$ at 310K.[95] Hence, the binding of dPGS to HSA resembles closely the one of poly(acrylic acid) to this protein discussed above.[91]

The analysis of the dependence of the binding constant $K_b$ and $\Delta G_b$ on temperature led to a result seen for many systems of biological relevance: $\Delta G_b$ hardly depends on temperature while $\Delta H_b$ and $\Delta S_b$ strongly vary with $T$.[76,123] The small dependence of $\Delta G_b$ on temperature inevitably leads to the conclusion that the specific heat of binding $\Delta C_p$ is of appreciable magnitude and that there is a strong enthalpy-entropy cancellation (EEC). Hence, the dependence of $\Delta G_{b,ITC}$ on temperature is given by the generalized van't Hoff equation (see e.g. ref.[82] and further references given there):

$$\Delta G_b = -RT \ln K_b = \Delta H_{vH,ref} - T\Delta S_{vH,ref} + \Delta C_{p,vH}\left[(T - T_{ref}) - T\ln\left(\frac{T}{T_{ref}}\right)\right] \quad (4)$$

where $\Delta H_{vH}$ and $\Delta S_{vH}$ and $\Delta C_{p,vH}$ are the respective thermodynamic functions derived from this analysis while $T_{ref}$ is a reference temperature.[95] All quantities deriving from $\Delta G_b$ refer directly to the binding equilibrium and thus give the respective enthalpy or entropy of binding. Figure 7 displays all data derived from this analysis.

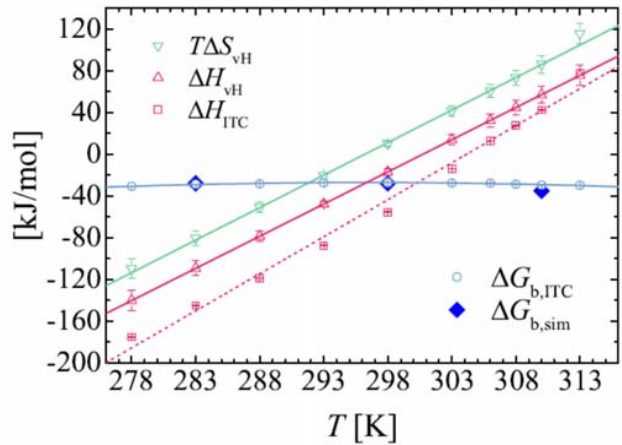

Figure 7 Dependence of the enthalpy of binding on temperature. The quantities $\Delta H_{vH}$ and $\Delta S_{vH}$ deriving from the analysis of $\Delta G_{b,ITC}$ on temperature are plotted as the function of temperature. $\Delta G_{b,ITC}$ and $\Delta G_{b,sim}$, on the other hand, hardly depend on temperature on this scale. The enthalpy $\Delta H_{ITC}$ measured directly by ITC differs from $\Delta H_{vH}$ that refers directly to the process of binding. The difference can be traced back to additional equilibria linked to the process of binding.[95]

The data displayed in Figure 7 are typical for the EEC seen in many binding equilibria in biological systems.[76,80–82] The present comparison, however, allows us to go one step beyond this much-discussed[76] result: The Gibbs free energy $\Delta G_b$ is shown here to result practically total from electrostatic effects which follows from the excellent agreement of $\Delta G_{b,ITC}$ and $\Delta G_{b,sim}$. Therefore we come to the conclusion that the EEC must be total for the present binding equilibrium. As a consequence, enthalpies of binding as well as $\Delta H_{ITC}$ measured directly by ITC does not tell us anything directly about the process of binding and all calculations must strive to obtain $\Delta G_{b,sim}$ (see also the discussion of this problem in ref.[76]).

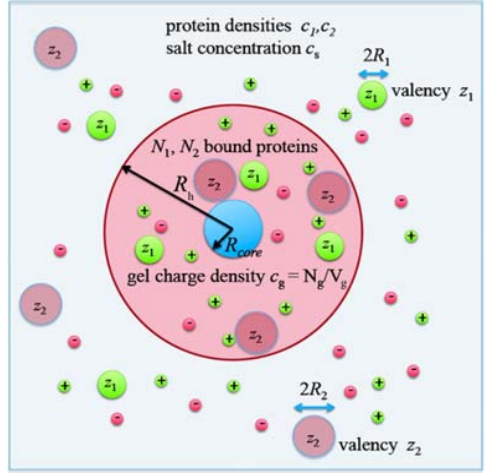

**Figure 8** Scheme of a core-shell microgel and the concentrations of the ions and the protein. The core is a solid sphere (marked in blue) onto which a weakly charged polymer gel is attached. The concentrations of co- and counterions in the gel are determined by the Donnan-equilibrium which also defines the leading term of the Gibbs free energy of protein adsorption. Taken from ref.[26].

## 3. PROTEINS INTERACTING WITH WEAKLY CHARGED GELS
### 3.1 Theory

Weakly charged polymer gels have been a long-standing subject in polymer science. We have studied the sorption of proteins to such a gel using the core-shell particles depicted schematically in Figure 8: A weakly charged gel is attached chemically to a solid core.[26] The notation 'weakly charged' means that the charged chains constituting the gel have a Manning parameter smaller than unity because the distance between two charges along the chains is larger than the Bjerrum length. Immersed in water, the gel layer will swell and proteins characterized by a net charge $z_p$ will be taken up. In general, the driving force can be split up into an electrostatic part $\Delta G_{el}$ and a non-electrostatic part that has been termed $\Delta G_0$ in ref[26]. The latter term will be operative also in non-charged gels and is governed by hydrophobic attraction. It turns out that the leading order contributions to the electrostatic term $\Delta G_{el}$ can be calculated in an analytical fashion:[26,28]

- The Donnan-potential $\Delta\phi$ can be calculated from the ratio of the charge density $c_g$ inside the gel and the salt concentration $c_s$ outside the gel. For an ideal solution we obtain $e\beta\Delta\phi = \ln[y + \sqrt{y^2 - 1}]$ where $y = z_g c_g/(2c_s)$ ($z_g$: valency of charge within network; $\beta = 1/k_B T$). Hence, the leading term for electrostatic attraction of the protein to the gel is given by $\Delta G_{el} = z_p \Delta\phi$. This leading term can be negative as well as positive depending on the sign of $z_p$. [The Donnan-equilibrium leads also to an increased osmotic pressure inside the gel due to the confined co- and counterions that can be calculated analytically, too (see eq.(2) of reference[26])].
- There is a second contribution to the electrostatic part that is always negative, that is, always attractive:[26] Proteins present charged entities that necessarily entail a Born self-energy of charging.

Thus, going from an uncharged sphere to a sphere with a net charge $z_p$ necessitates a positive free energy due to the repulsive interaction among the charges. Repeating the same process in a medium with higher ionic strength leads to a smaller Born energy due to screening. Hence, the charged proteins tend to go from a solution with low ionic strength into the gel in which the salt concentration is higher.

We obtain for the term related to the Born energy up to the monopole term the following expression[26]

$$\beta \Delta G_{Born} = \frac{z_p^2}{2R_p}\left(\frac{\kappa_g R_p}{1+\kappa_g R_p} - \frac{\kappa_b R_p}{1+\kappa_b R_p}\right) \quad (5)$$

where the inverse screening lengths are $\kappa_g \approx (8\pi l_B c_g)^{1/2}$ and $\kappa_b = (8\pi l_B c_s)^{1/2}$ for gel and bulk, respectively. For a patchy surface of a typical protein (see Figure 1), the calculation of the Born energy becomes more difficult but an approximation on the dipolar level can be obtained.[93]

- Uptake of proteins with a net charge $z_p$ must change the charge density $c_g$ of the gel. For instances, if the network is charged negatively and the protein carries along a net positive charge, the charge density within the gel decreases which must be taken into account when calculating $\Delta\phi$ (see the discussion of eq.(4) of ref.[26]). This effect must also be taken into account when calculating the osmotic pressure within the gel and the swelling equilibrium.

Given these prerequisites and the terms, the electrostatic part $\Delta G_{el}$ is fully determined. It must be kept in mind that $\Delta G_{el}$ is an explicit function of the number of proteins taken up by the gel. Hence, the first protein will experience a much stronger electrostatic interaction than the proteins taken up near to saturation. In this way, the uptake of proteins to charged gels experience a negative cooperativity exactly in the way as discussed above for the case of charged dendrimers.

Proteins immersed in a gel will exhibit a repulsive mutual interaction that increases with increasing volume fraction. Since the ionic strength within the gel is rather high, this repulsion can be modeled to good approximation by the hard sphere interaction. Thus, the repulsive chemical potential of the proteins within the gel can be approximated by the Carnahan-Starling (CS) excess chemical potential:[128]

$$\beta \mu_{CS} = \frac{8\eta - 9\eta^2 + 3\eta^3}{(1-\eta)^3} \quad (6)$$

which is the free energy necessary for transferring one hard sphere into a volume with a packing fraction $\eta = (N_{bp}/V_{gel})\pi\sigma_P^3/6$ and $\beta = 1/k_BT$. Here $N_{bp}$ is the number of polymers in the gel, $V_{gel}$ is the actual volume of the gel available for the proteins and $\sigma_p$ is the diameter of the protein which is approximated by a sphere. Evidently, $V_{gel}$ must be corrected for the volume fraction of the polymer and for the shrinking of the gel with an increasing number of proteins taken up. Given these various terms, the total number of proteins can be formulated as

$$\frac{N_{bp}}{V_{gel}} = \zeta' c_p \exp[-\beta \Delta G_{el} - \beta \Delta G_0] \exp[-\beta \mu_{CS}] \quad (7)$$

where $\zeta'$ is the partition function of the bound state, i.e., considers constraints of protein degrees of freedom such as rotation or vibration in the bound state. In first approximation, the proteins can be regarded as free entities with only a translational degree of freedom that may move around freely in the network, *apart from the separately considered packing effects,* and $\zeta'=1$. This model was named "Excluded Volume (EV) model[26] as the binding is limited at large coverage by protein packing. Ideally $\Delta G_0$ describes only the non-electrostatic contributions and thus should not depend on ionic strength but only on temperature. In the following it will be treated as an adjustable parameter.

It is revealing to compare the EV-model thus defined to the conventional Langmuir model used routinely to analyze the adsorption of proteins to gels and to spherical polyelectrolyte brushes.[24,25,27] We could demonstrate that the EV model reduces to the Langmuir isotherm if the packing fraction is rather low, that is, if $\mu_{CS}$ can be approximated in terms of the second virial coefficient $B_2 = 2\pi\sigma_p^3/3$ of a system of hard spheres.[26] $2B_2$ is the volume excluded by a single sphere for all other spheres in this approximation and it turns out that the Langmuir isotherm refers to a system in which the proteins are located in $N_{bp}$ binding boxes each of which has a volume of $2B_2$. Thus, the maximum number of proteins to be bound to a gel follows from the maximum number of places given by $V_{gel}/(2B_2)$. It should be noted that the second virial coefficients can be obtained directly from scattering experiments or calculated from crystallographic data.

The above correspondence between the EV-model and the Langmuir isotherm can now be used to extend the calculation to competitive adsorption of several proteins.[28] If the packing fraction $\eta$ of all proteins is not too high, the chemical potential $\mu_{CS,i}$ of the $i$th protein can be approximated by

$$ß\mu_{CS,i} = 2 \sum_j B_2^{ij} \frac{N_j}{V_{gel}} \qquad (8)$$

where the second virial coefficient $B_2^{ij}$ for a pair of interacting proteins $i$ and $j$ can be written down exactly through

$$B_2^{ij} = \frac{2}{3}\pi \left[\frac{\sigma_i+\sigma_j}{2}\right]^3 \qquad (9)$$

if the proteins can be modeled as hard spheres in sufficient approximation. Thus, the repulsive term can be generalized to an arbitrary number of proteins. The same holds true for the electrostatic terms and $\Delta G_0$ is referring to the interaction of a given single protein to the network and should be independent of the presence of other proteins for small packing fraction. Thus, the entire treatment can be used to consider the competitive adsorption of a mixture of many proteins to a gel in equilibrium. This problem which is far more practical than the adsorption of a single protein is hence within reach of a quantitative modeling.

**a)**

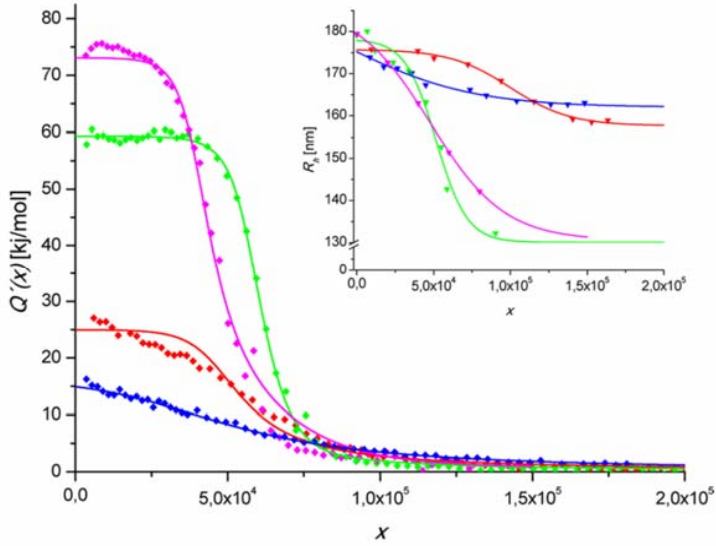

**b)**

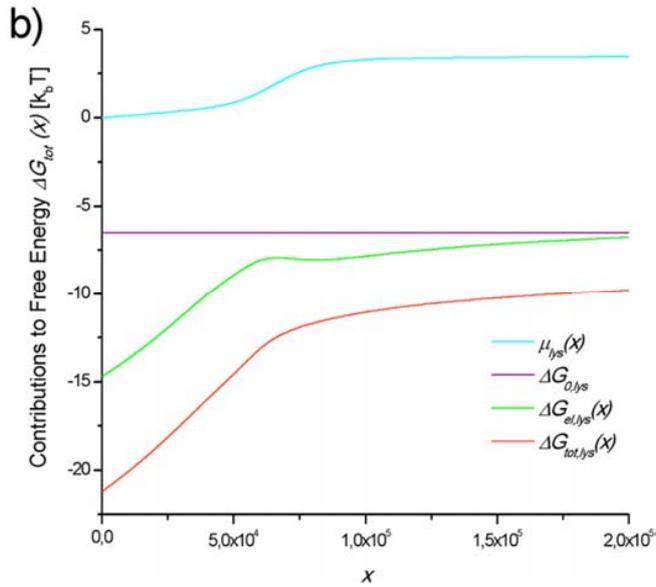

**Figure 9** Experimental observation of the adsorption of proteins to core-shell microgels by ITC.[28] **a)** Adsorption isotherms for four different proteins measured at 7mM ionic strength and 298K. The differential heat is plotted against the molar ratio $x$ of the enzymes to the particles. The enzymes are: papain (magenta), lysozyme (green), cytochrome C (red), RNase A (blue). The solid lines denotes the fits with eq.(6). The inset shows the hydrodynamic radius of the microgels as measured by dynamic light scattering. The shrinking with the uptake of protein is clearly visible. The solid lines display a fit with an empirical model. **b)** The total Gibbs free energy of binding for lysozyme is decomposed into its components. $\Delta G_0$ denotes the intrinsic (non-electrostatic) adsorption free energy; $\Delta G_{el}(x)$ is the electrostatic contribution; $\mu_{lys}(x)$ is the entropic penalty due to hard sphere packing calculated by the Carnahan−Starling potential.

*3.2 Experiment*

In the following we shall review the experiments done on charged core-shell microgels.[24–26,28] The experimental methods applied in these studies have been discussed in earlier reviews[27,129] so that the present review can be focused on the comparison of the experimental data with theory. Eq.(7) gives the dependence of concentration of proteins in a given volume of the gel $V_{gel}$ as the function of the total free energy of binding. The comparison with the ITC-experiments can be done as follows: ITC measures the heat flux per injection of a dilute solution of protein to a solution of the microgel particles. Figure 9a displays the experimental data taken for four different proteins at a low ionic strength.[28] Here the heat of adsorption per injection $Q'$ is plotted against the molar ratio $x$ of the respective protein to the microparticle. Strong adsorption leads to a plateau at small x that directly gives the heat of adsorption. $\Delta H_{ITC}$ determined by ITC. Larger $x$ are followed by a decrease of $Q'$ until saturation is achieved. These data can be mapped in the usual way[26] on a Langmuir isotherm with a constant $K(x)$ being an explicit function of the molar ratio $x$. Thus, we have

$$K(x) = K_0 \exp[-\beta \Delta G_{el}(x)] = \frac{\theta(x)}{[1-\theta(x)]c_p} \qquad (10)$$

This fit takes into account the explicit dependence of the Gibbs free energy of the number of proteins already taken up by the gel. In this way the negative cooperativity of binding due to electrostatics is taken care of. Moreover, the marked shrinking of the gel displayed in the inset of Figure 9a is taken into account so that the actual volume $V_{gel}$ is used at each $x$.

Figure 9b displays the different terms of the Gibbs free energy of adsorption. Note that only $\Delta G_0$ is an adjustable parameter, all other terms can be calculated from the number of adsorbed proteins. In principle, the non-electrostatic part $\Delta G_0$ should be independent of the ionic strength in the system. The data obtained in ref.[26] show that this is the case in rather good approximation and $\Delta G_0$ is found to increase only slightly with $c_s$. Given the various approximations and assumption in this fit, this result can be viewed upon as full agreement of theory and experiment. Figure 9b demonstrates that all other contributions vary strongly with $x$. The electrostatic contribution $\Delta G_{el}$ is strongly negative at low $x$ but its magnitude decreases with increasing uptake of protein due to the decrease of the overall charge of the microgel. The repulsive part $\mu_{CS}$ is positive, of course, and becomes more important only at higher $x$. Evidently, fitting the set of data to a single binding constant $K$ which is possible would inflict a grave error in the assessment of the Gibbs free energy driving the uptake of protein.

The value of $\Delta G_0$ can thus be obtained for different proteins can in turn be used to compare the competitive adsorption of two proteins to a given microgel.[28] This process can be investigated by using lysozyme bearing a fluorescent label. Earlier work has shown that the fluorescence of the label is quenched as soon as the labeled lysozyme is taken up by the gel.[24] This effect is due to the slightly smaller pH within the gel and can be used to monitor the replacement of labeled lysozyme by another protein. The freed lysozyme fluoresces again and its concentration can be measured quantitatively as the function of the amount of added protein. In our experiments we first validated this procedure by replacing labeled lysozyme by unlabeled lysozyme. Moreover, the competitive adsorption of lysozyme with cytochrome C, Rnase A, and papain was determined by the same method. The concentration of lysozyme set free by the addition of these proteins can be calculated without adjustable parameters through the use of eq.(6). Here full agreement of theory and experiment was found for unlabeled lysozyme, cytochrome C, and RNase A, only in case of papain we found small deviation between the predicted and the measured amount of freed lysozyme.

Thus, all comparisons of theory and experiment done on charged microgels have met with gratifying success so far. This good agreement can be taken as a proof that the Gibbs free energy of adsorption of proteins to charged gels can be calculated and modeled in very good approximation. It should be noted, however, that this comparison of ITC to theory has tacitly assumed that the measured heat of adsorption is solely due to the interaction with the charged gel and not due to any distortion of the tertiary or secondary structure. We could show that the native structure of lysozyme is not changed by the uptake into the charged gel.[25] This could be demonstrated directly by measuring the enzymatic activity of the bound lysozyme. Here we found an even elevated activity as compared to the free enzyme which could be explained by the slightly lower pH within the gel. It is hence evident that the tertiary structure has been fully preserved within the network which is expected from the rather weak interaction of the proteins with the slightly charged network. Partial unfolding of proteins, however, may become a serious issue for carrier system interacting more strongly with proteins. In this case a check of enzymatic activity[97] or of the secondary structure by FT-IR[130] are very helpful.

4. PROTEINS INTERACTING WITH SPHERICAL POLYELECTROLYTE BRUSHES

If linear polyelectrolytes are attached densely by one end to a planar or curve surface they form a planar polyelectrolyte brush or a spherical polyelectrolyte brush (SPB). These systems have been well-studied during the last decade and their general behavior is well-understood.[131] The main feature of polyelectrolyte brushes is the strong confinement of the counterions. Similar to the charged networks discussed further above the brushes will be swollen by the uptake of water in order to release the osmotic pressure of the confined counterions. Hence, in salt-free solutions an *osmotic limit* is reached in which the chains of the brush layer are strongly stretched. In the *salted limit*, that is, at high salt concentrations the electrostatic interactions are strongly screened and the chains of the brush layer assume a configuration as in a uncharged brush. Some time ago it was found that proteins will be adsorbed on a brush layer despite the fact that both the protein and the brush bear the same net charge (adsorption on the "wrong side" of the isoelectric point) if the brush is in the osmotic limit.[47] If the ionic strength is raised, there is only a weak or no adsorption. Protein adsorbed in the osmotic limit will be released upon raising the ionic strength.[132] This phenomenon has been shown to be general and a survey of the literature has been given in a number of reviews.[27,90,48,133] A comprehensive theoretical treatment has been presented recently[93] and the following section will summarize the salient points of this analysis. In particular, uptake of proteins on polyelectrolyte brushes can now be compared with the well-studied case of charge networks reviewed in section 3 above.

## 4.1 Theory

In principle, proteins should be repelled by a brush of like charge because of the electrostatic and steric repulsion: Insertion of a protein into a brush layer will lead to steric interactions with the densely tethered chains and increase the osmotic pressure of the confined counterions. These adverse effects can be overcome by three major forces:

i) The pH value within the brush layer can be considerably lower as outside in the bulk solution due to confined protons. If this pH is lower than the isoelectric point of the protein, the charge of the immersed protein will be change to a positive value leading to a strong attraction. [134]

ii) Counterion release was evoked early on as a strong attractive force.[47] Figure 10a displays the main feature of this attraction: Positive patches of the protein become multivalent counterions of the brush layer thus releasing a concomitant number of counterions confined within the brush layer. This force already discussed above for the interaction of proteins with single chains of polyelectrolytes will be even stronger due to the nearly 100% confinement of the ions within the brush layer. [27,48,90,93,133]

iii) Proteins can have large dipole moments because of the surface patches of charge. In case of spherical polyelectrolyte brushes there is a non-vanishing electric field in the brush layer the interaction of which with the proteins will lead to a marked attraction.[93]

These considerations have been put into a more quantitative treatment of the uptake of proteins to polyelectrolyte brushes.[93] First, MD-simulations have been done using a coarse-grained model of a planar polyelectrolyte brush whereas the proteins have been modeled by the charged patchy protein model (CPPM) introduced in section 2 above (see Figure 1). The binding of a protein could again be studied by a steered Langevin-simulation which leads to the Gibbs free energy of binding. In this way results obtained on polyelectrolyte brushes can be compared directly to results obtained from simulations of proteins with single polyelectrolyte chains (see section 2) and with charged networks (see section 3). In a second step, these simulations could be compared to approximate analytical expressions of the various terms contributing to the free energy of binding:[93]

- The first contribution to $\Delta G_b$ is related to the immersion of a neutral globule into a charged brush. This term is dominated by the steric repulsion between the neutral globule and the brush due to the steric interaction of the globule with the chains and by the raise of the osmotic pressure of the confined counterions. Estimates of both contributions show clearly that the raise of the osmotic pressure is dominating and may be of the order of several tens of $k_BT$.

- Next, the electrostatic contributions can be discussed, much in a way already worked out for the charged gels (see section 2). The concentration of counter- and coions within the brush layer can be well-approximated by a Donnan equilibrium. The resulting *Donnan-potential* defines the monopole term which is positive for the uptake of a negatively charged protein to a brush layer bearing the like charge. Hence, the term corresponds directly to the Donnan-term already formulated for the case of charged gels. In addition to this, there is the interaction of the strong dipole of the protein with the electric field of the brush layer. In case of the rather homogeneous charged gels this term can be omitted in first approximation. It is operative in the case of the spatial inhomogeneous brush layers and cannot be neglected for spherical polyelectrolyte brushes.

- The *Born term* already introduced when discussing the uptake of proteins in gels must be considered here, too, since the salt concentration inside the brush layer differs considerably from the salt concentration in bulk. For the present systems it is always negative. In ref.[93] we developed this term up to the dipole level which can be done analytically for point dipoles in the Debye-Hückel approximation (see eq.(6) of ref.[93]). All expressions refer to homogenous systems but can be applied in to the inhomogeneous systems under consideration here as well. However, a comparison with the MD-simulations suggests that the Born term as presented in ref.[93] is the weakest approximation of all analytical expressions.
- Finally, counterion release could be modeled as already outlined for single polyelectrolyte chains in section 2.

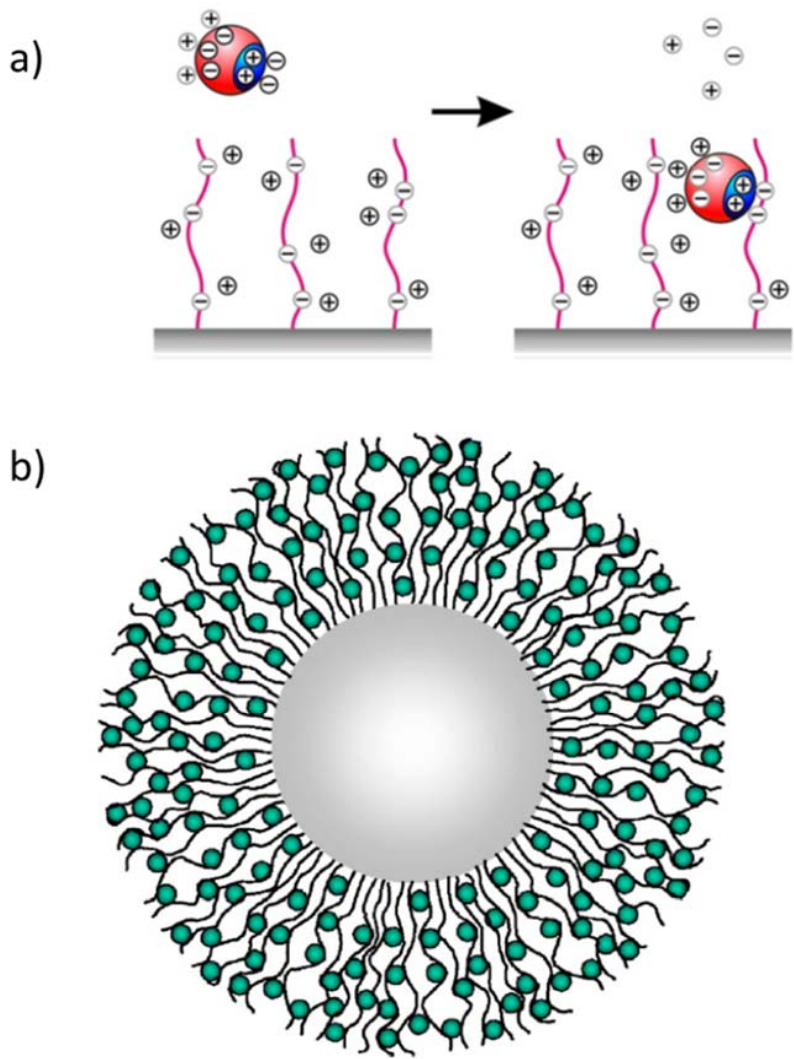

**Figure 10** Uptake of proteins by polyelectrolyte brushes.[83,84] **a)** Counterion release mechanism for the binding of proteins to brush layers. A positively charged patch on a protein can replace the counterions in a negatively charged polymer brush despite the fact that the overall charge of the protein is negative. The counterions from the surface of the protein, and the counterions from the polymer brush are released into the bulk of the solution. The increase in entropy upon counterion release is the main driving force for protein adsorption on the "wrong side" of the isoelectric point, i.e. when the net charge of the protein globule has the same sign as the chains in the brush.[48,135] **b)** Spatial structure of a complex between proteins marked in green with a spherical polyelectrolyte brush.[83,84] The proteins are closely correlated to the polyelectrolyte chains of the brush layer.

With these analytical expressions in place, the various contributions to the Gibbs free energy of binding could be compared to the MD-simulations. Here it turned out that the final negative Gibbs free energy of binding is the result of cancelation of several large terms differing in sign: The monopole repulsion can go up to 23 $k_BT$ whereas the Born term cancels this contribution largely with a magnitude up to -18 $k_BT$. The dipolar contribution is largest at the surface of the brush as expected but remains negative within the brush. Counterion release is strongly attractive with $10 - 14$ $k_BT$ depending on salt concentration. It scales very well with $\ln(c_s)$ as predicted by eq.(1). The resulting Gibbs free energy is overall negative and leads to a strong binding. The analytical expression for $\Delta G_b$, however, underestimates considerably the value resulting from MD-simulations, most probably due to problems in the approximations involved in calculating the Born term (which essentially reflect neglecting higher order terms in a multipole expansion of the electrostatic potential).

*4.2 Experiments*

Up to now, there is no direct comparison of theory and experiment. However, all experimental results obtained so far corroborate the conclusions of theory at least in a semi-quantitative fashion. First of all, the decisive parameter governing the interaction of proteins with polyelectrolyte brush layers is the ionic strength in the system. This is shown schematically in Figure 11 displaying the amount of adsorbed protein per gram of the SPB $\tau_{ads}$ as the function of the concentration $c_{sol}$ of the protein remaining in solution. In this way the resulting curves can be compared to typical adsorption isotherms. Only at high ionic strength the brush layer becomes more and more protein-resistant because of the steric repulsion between the dissolved proteins and the brush layer of the SPB.

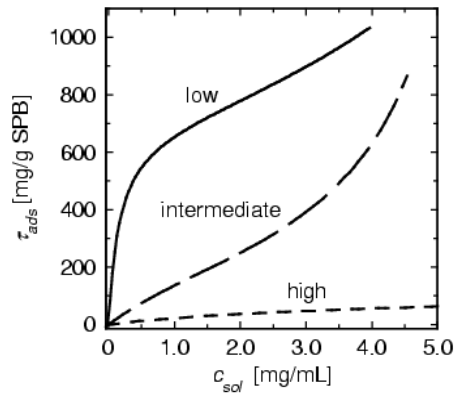

**Figure 11** Schematic representation of the adsorption of proteins onto SPBs. The amount of adsorbed protein per gram of the carrier particles $\tau_{ads}$ is plotted against the concentration of the protein $c_{sol}$ remaining in solution. Parameter of the curves is the concentration $c_s$ of added salt defining the ionic strength in the system. Strong adsorption takes place at low ionic strength whereas little protein is adsorbed if the ionic strength is high.[48]

The entropic origin of the strong adsorption of proteins to polyelectrolyte brushes could be directly demonstrated by isothermal titration calorimetry (ITC).[98] A marked endothermic signal was observed when titrating β-lactoglobulin into a solution of spherical polyelectrolyte brushes bearing polystyrene sulfonate chains. This entropy could directly be used to estimate the number of released counterions by eq.(1). Approximately 10 counterions in total were released upon uptake of a single protein molecule. This binding entropy decreased significantly when raising the ionic strength in the system as expected for the counterion release force.[98,136] Studies by small-angle X-ray scattering furthermore demonstrated that the proteins are closely bound to the polyelectrolyte chains of the brush layer (see the scheme shown in Figure 10b).[83–86] Similar results have been obtained on planar systems by neutron reflectometry.[137] Additional analysis by FT-IR spectrocopy[84,130,138] and by checks of the enzymatic activity of bound enzymes[97,139,140] demonstrated clearly that the secondary and the tertiary structure of the bound proteins is not changed for brush layer bearing the hydrophilic chains of poly(acrylic acid). This finding was corroborated by fluorescence studies of the green fluorescent protein.[88]

*4.2 Dynamics of the uptake of proteins to microgels and spherical polyelectrolyte brushes*

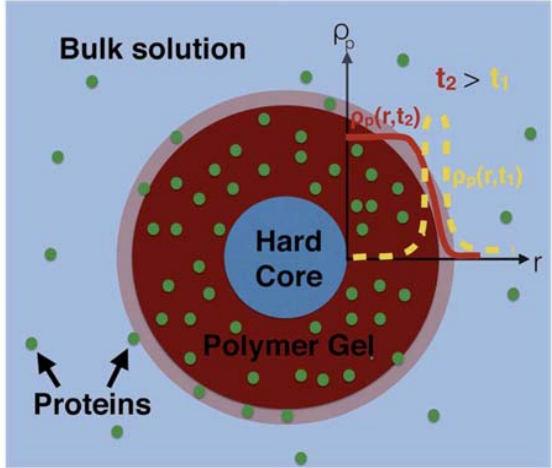

**Figure 12** Dynamics of the uptake of proteins to charged microgels.[92] The core-shell particle consists of a hard core (blue) and a shell consisting of a cross-linked polymer network (red) that has an interface with a width of 10 nm (light red). The proteins are represented by green spheres and the particles are immersed in a protein solution. For the symmetry of this system, the protein concentration is modeled as a radial density field $\rho_p(r,t)$ with the origin being located at the nanoparticle's hard core/polymer boundary. In Figure 12 two density profiles corresponding to different times are shown: yellow dashed line referring to earlier time and red referring to a later time.

In the following a brief summary of recent work on the dynamics of the uptake of proteins by microgels and spherical polyelectrolyte brushes will be given. The exposition given here follows the rendition in refs.[92,105] In sections 3 and 4 we have demonstrated that the adsorption of proteins to complex particles can be described in terms of a free energy that depends on the local number density of the proteins $\rho$. In formulating our model, we have implicitly assumed the presence of two states only. One, where the protein is adsorbed in the bulk of the microgel, the other where the protein is in the bulk of the solution. Clearly, this is a simplification because $\rho_p$ will vary *continuously* between these two limits as the local environment interpolates between that of a bulk gel and that of a bulk solution. The full spatial dependence of $\rho_p$ can be calculated via (classical) Density Functional Theory[141]. In practice, DFT states that the local chemical potential $\mu_p(r)$ is a function of the density field $\rho_p(r)$ and is given by the functional derivative:

$$\mu_p = \frac{\delta F[\{\rho_p\}]}{\delta \rho_p} \qquad (10)$$

where $F[\{\rho_p\}]$ is the free-energy *functional* of the system under consideration and depends on the density $\rho_p$ of *all* species. In equilibrium, the chemical potential of each protein type is constant throughout the system and fixing its values provides a self-consistent equation to find $\rho_p(r)$. In Ref.[92,105] we have shown how to construct a free-energy functional that reproduces the two-state model previously described in Sec. 3.1 when the bulk states are approached.

One of the main advantages of a DFT reformulation of the protein adsorption problem is to provide a thermodynamically consistent way to study the *dynamics* of adsorption including the effect of protein-protein and protein-gel interactions. For doing this, we need to make two further assumptions. The first is that proteins follow a relaxation dynamics, e.g., Brownian dynamics[142]. The second is a separation of time-scales similar to the Born-Oppenheimer approximation: we assume that the degree of freedom of all other species, e.g., solvent molecules and ions, instantaneously relax to their local equilibrium configuration given a fixed protein density field.[105] In other words, we require proteins to be much slower than all other species in the system. With these prerequisites, broadly satisfied under normal experimental conditions, the Dynamic Density Functional Theory (DDFT) allows to recast the time evolution of the protein density field into a Generalised Diffusion Equation:[92]

$$\frac{\partial \rho_p}{\partial t} = \nabla D_p \rho_p \nabla \beta \mu_p \qquad (11)$$

where $D_p$ denotes the diffusion coefficient of species $p$. Note that in Eq. 11 the chemical potential is a function of all the density field $\{\rho_p\}$ for all protein types $p$ present, as described by Eq. 10. Hence, Eq.11 represents a non-linear, partial differential equation for $\rho_p$. If the chemical potential $\mu_p$ is given by the ideal solution term, i.e. no interactions are considered, we recover what is simply known as the diffusion equation. In this case, the system will simply try to smoothen every possible gradient. However, when thermodynamic interactions are included, the picture becomes much more complex.

Figure 12 shows the application of this DDFT model to the adsorption of proteins to charged microgels (see section 3). In ref.[92], we have shown that, unlike the simple diffusion equation, a DDFT-based model can reproduce semi-quantitatively the dynamics of protein adsorption as measured by fluorescence experiments.[24] More recently, we have applied the same formalism to describe the dynamics of a mixture of proteins.[105] In particular, we have shown how the often observed non-monotonic behavior in the adsorption of proteins, commonly known as the "Vroman effect"[143–147], does not have a single origin but can generally arise from completely different mechanisms involving two or more competing interactions. (see Figure 13). The general take-home message from these simulations is that trends on the adsorption kinetics and thermodynamics derived from single-protein type experiments cannot be straightforwardly extrapolated to mixtures. For example, evaluating the performance of anti-adsorption polymer coatings based on such experiments might lead to qualitatively wrong conclusions, because protein-protein interactions, either direct or mediated through global electrostatics, can drastically change the adsorption behavior.

For example, the same formalism applied to describe the Vroman effect also predicts that for certain protein mixtures co-adsorption in the microgel can occur. In this case, a protein can increase its adsorption by almost two orders of magnitude when in a mixture with other proteins, compared to the case where the protein is present alone. Quite interestingly, our simulations also highlight the importance of describing the whole spatial variation of the protein density field rather than simply looking at the bulk states. In particular, we have observed how in some cases the composition of the bulk of the gel is not just quantitatively but also qualitatively different from that of the gel-solution interface. This is of particular importance considering that it is this outer layer of proteins that determines the interaction of the microgel with its surrounding, e.g. the immune system. In this case, similarly to what happens for single vs multiple protein mixtures, interpreting experimental data based on the overall amount of protein adsorbed, dominated by the bulk of the gel, can again lead to completely wrong conclusions.

It is interesting to note that a simplified model along the same lines used to describe protein adsorption on microgels has been used to study the uptake of proteins also to spherical polyelectrolyte brushes.[85] The radial density of the chains decays with radial distance of the core that leads to a concomitant decay of the overall attractive potential that is directly coupled to the segmental density. Assuming a frictional coefficient of the protein molecules, this simple model leads to the conclusion that the overall amount of proteins scales with $t^{1/4}$ which is observed experimentally by time-resolved small-angle x-ray scattering.

In general, the message arising from these studies is that inclusion of thermodynamic interactions is crucial to describe the dynamics of protein adsorption in these polymeric carriers. Since a different model can be built by simply changing the free-energy functional, we expect this formalism to lead to predict new and interesting phenomena in other gel-protein systems. Furthermore, it should be noticed that in the same way not only protein adsorption but also protein desorption could be addressed and work along this line is already ongoing.

## 5. CONCLUSION

We reviewed recent work on the interaction of proteins with various polyelectrolytes with special emphasis on the quantitative comparison of theory and experiment. Two major strands of systems can be distinguished: i) single chains of polyelectrolytes or hyperbranched/dendritic polymers, and ii) systems of polyelectrolyte as charged networks and polyelectrolyte brushes. Counterion release was found to be the major driving force for protein binding except for the case of weakly charged gels. Moreover, in all cases approximate expressions could be derived to model protein adsorption. In case of multiple adsorption of proteins, a negative cooperativity of adsorption could be clearly seen: The magnitude of the Gibbs free energy of binding decreases considerably with the number of already bound proteins. The Gibbs free energy obtained in this way for a given system can then be used to understand and even predict the dynamics of protein uptake by this system as discussed in section 4.

Perhaps the most important point of all studies presented herein is the excellent agreement of simulated data and theoretical modeling with experimental binding constants. We see this agreement for very hydrophilic single and dendritic polyelectrolytes where no adjustable parameter is necessary in coarse-grained simulations. In case of charged gels a single adjustable parameter describing non-electrostatic contributions had to be introduced. These surprising findings suggest that the binding of proteins to hydrophilic polyelectrolytes is governed by electrostatic factors as counterion release and Debye-Hückel interaction for low ionic strength, no terms related to hydration intervene. The consequences of this finding remain to be explored in more detail for various applications as e.g. drug design and for investigations of systems of direct biological relevance.[62,63]

ACKNOWLEDGMENT X.X. thanks the Chinese Scholar Council for financial support. S. A.-U. and J. D. acknowledge support by the Alexander-von-Humboldt-Foundation. The International Research Training Group IRTG 1524 funded by the German Science Foundation and the Helmholtz Virtual

Institute for Multifunctional Biomaterials for Medicine are gratefully acknowledged for financial support.Institute for Multifunctional Biomaterials for Medicine are gratefully acknowledged for financial support.

REFERENCES

(1) De Kruif, C. G.; Weinbreck, F.; De Vries, R. Complex Coacervation of Proteins and Anionic Polysaccharides. *Curr. Opin. Colloid Interface Sci.* **2004**, *9*, 340–349.

(2) Kizilay, E.; Kayitmazer, A. B.; Dubin, P. L. Complexation and Coacervation of Polyelectrolytes with Oppositely Charged Colloids. *Adv. Colloid Interface Sci.* **2011**, *167* (1–2), 24–37.

(3) Xu, Y.; Liu, M.; Faisal, M.; Si, Y.; Guo, Y. Selective Protein Complexation and Coacervation by Polyelectrolytes. *Advances in Colloid and Interface Science*. January 2017, pp 158–167.

(4) Park, J. M.; Muhoberac, B. B.; Dubin, P. L.; Xia, J. Effects of Protein Charge Heterogeneity in Protein-Polyelectrolyte Complexation. *Macromolecules* **1992**, *25* (1), 290–295.

(5) Xia, J.; Dubin, P. L.; Kim, Y.; Muhoberac, B. B.; Klimkowski, V. J. Electrophoretic and Quasi-Elastic Light Scattering of Soluble Protein-Polyelectrolyte Complexes. *J. Phys. Chem.* **1993**, *97* (17), 4528–4534.

(6) Gao, J. Y.; Dubin, P. L.; Muhoberac, B. B. Measurement of the Binding of Proteins to Polyelectrolytes by Frontal Analysis Continuous Capillary Electrophoresis. *Anal. Chem.* **1997**, *69* (15), 2945–2951.

(7) Grymonpré, K. R.; Staggemeier, B. A.; Dubin, P. L.; Mattison, K. W. Identification by Integrated Computer Modeling and Light Scattering Studies of an Electrostatic Serum Albumin-Hyaluronic Acid Binding Site. *Biomacromolecules* **2001**, *2* (2), 422–429.

(8) Seyrek, E.; Dubin, P. L.; Tribet, C.; Gamble, E. A. Ionic Strength Dependence of Protein-Polyelectrolyte Interactions. *Biomacromolecules 4* (2), 273–282.

(9) Cooper, C. L.; Dubin, P. L.; Kayitmazer, A. B.; Turksen, S. Polyelectrolyte–protein Complexes. *Curr. Opin. Colloid Interface Sci.* **2005**, *10* (1–2), 52–78.

(10) Xu, Y.; Mazzawi, M.; Chen, K.; Sun, L.; Dubin, P. L. Protein Purification by Polyelectrolyte Coacervation: Influence of Protein Charge Anisotropy on Selectivity. *Biomacromolecules* **2011**, *12* (5), 1512–1522.

(11) Cooper, C. L.; Goulding, A.; Kayitmazer, a. B.; Ulrich, S.; Stoll, S.; Turksen, S.; Yusa, S. I.; Kumar, A.; Dubin, P. L. Effects of Polyelectrolyte Chain Stiffness, Charge Mobility, and Charge Sequences on Binding to Proteins and Micelles. *Biomacromolecules* **2006**, *7*, 1025–1035.

(12) Kayitmazer, a. B.; Quinn, B.; Kimura, K.; Ryan, G. L.; Tate, A. J.; Pink, D. a.; Dubin, P. L. Protein Specificity of Charged Sequences in Polyanions and Heparins. *Biomacromolecules* **2010**, *11*, 3325–3331.

(13) Kayitmazer, a. B.; Seeman, D.; Minsky, B. B.; Dubin, P. L.; Xu, Y. Protein–polyelectrolyte Interactions. *Soft Matter* **2013**, *9*, 2553.

(14) Comert, F.; Malanowski, A. J.; Azarikia, F.; Dubin, P. L. Coacervation and Precipitation in Polysaccharide-Protein Systems. *Soft Matter* **2016**, *12* (18), 4154–4161.

(15) Lindhoud, S.; Voorhaar, L.; Vries, R. De; Schweins, R.; Stuart, M. a C.; Norde, W. Salt-Induced Disintegration of Lysozyme-Containing Polyelectrolyte Complex Micelles. *Langmuir* **2009**, *25* (14), 11425–11430.

(16) Weinbreck, F.; de Vries, R.; Schrooyen, P.; de Kruif, C. G. Complex Coacervation of Whey Proteins and Gum Arabic. *Biomacromolecules* **2003**, *4* (2), 293–303.

(17) de Vries, R.; Weinbreck, F.; de Kruif, C. G. Theory of Polyelectrolyte Adsorption on Heterogeneously Charged Surfaces Applied to Soluble Protein–polyelectrolyte Complexes. *J. Chem. Phys.* **2003**, *118* (10), 4649.

(18) de Vries, R.; Cohen Stuart, M. Theory and Simulations of Macroion Complexation. *Curr. Opin. Colloid Interface Sci.* **2006**, *11* (5), 295–301.

(19) KABANOV, V. PHYSICOCHEMICAL BASIS AND THE PROSPECTS OF USING SOLUBLE INTERPOLYELECTROLYTE COMPLEXES. *Vysokomol. Soedin. SERIYA A SERIYA B* **1994**, *36* (2), 183–197.

(20) Zezin, A.; Rogacheva, V.; Skobeleva, V.; Kabanov, V. Controlled Uptake and Release of Proteins by Polyelectrolyte Gels. *Polym. Adv. Technol.* **2002**, *13* (10–12), 919–925.

(21) Kabanov, V. A.; Skobeleva, V. B.; Rogacheva, V. B.; Zezin, A. B. Sorption of Proteins by Slightly Cross-Linked Polyelectrolyte Hydrogels: Kinetics and Mechanism. *J. Phys. Chem. B* **2004**, *108* (4), 1485–1490.

(22) Oh, K. T.; Bronich, T. K.; Kabanov, V. A.; Kabanov, A. V. Block Polyelectrolyte Networks from Poly(acrylic Acid) and Poly(ethylene Oxide): Sorption and Release of Cytochrome C. *Biomacromolecules* **2007**, *8* (2), 490–497.

(23) Mattison, K. W.; Dubin, P. L.; Brittain, I. J. Complex Formation between Bovine Serum Albumin and Strong Polyelectrolytes: Effect of Polymer Charge Density. *J. Phys. Chem. B* **1998**, *102* (19), 3830–3836.

(24) Welsch, N.; Dzubiella, J.; Graebert, A.; Ballauff, M. Protein Binding to Soft Polymeric Layers: A Quantitative Study by Fluorescence Spectroscopy. *Soft Matter* **2012**, *8* (48), 12043.

(25) Welsch, N.; Becker, A. L.; Dzubiella, J.; Ballauff, M. Core–shell Microgels as "smart" Carriers for Enzymes. *Soft Matter* **2012**, *8*, 1428.

(26) Yigit, C.; Welsch, N.; Ballauff, M.; Dzubiella, J. Protein Sorption to Charged Microgels: Characterizing Binding Isotherms and Driving Forces. *Langmuir* **2012**, *28* (40), 14373–14385.

(27) Welsch, N.; Lu, Y.; Dzubiella, J.; Ballauff, M. Adsorption of Proteins to Functional Polymeric Nanoparticles. *Polymer (Guildf)*. **2013**, *54* (12), 2835–2849.

(28) Oberle, M.; Yigit, C.; Angioletti-Uberti, S.; Dzubiella, J.; Ballauff, M. Competitive Protein Adsorption to Soft Polymeric Layers: Binary Mixtures and Comparison to Theory. *J. Phys. Chem. B* **2015**, *119* (7), 3250–3258.

(29) Li, Y.; de Vries, R.; Kleijn, M.; Slaghek, T.; Timmermans, J.; Stuart, M. C.; Norde, W. Lysozyme Uptake by Oxidized Starch Polymer Microgels. *Biomacromolecules* **2010**, *11* (7), 1754–1762.

(30) Lassen, B.; Malmsten, M. Competitive Protein Adsorption at Plasma Polymer Surfaces. *J. Colloid Interface Sci.* **1997**, *186*, 9–16.

(31) Johansson, C.; Hansson, P.; Malmsten, M. Mechanism of Lysozyme Uptake in Poly(acrylic Acid) Microgels. *J. Phys. Chem. B* **2009**, *113* (18), 6183–6193.

(32) Malmsten, M.; Bysell, H.; Hansson, P. Biomacromolecules in Microgels — Opportunities and Challenges for Drug Delivery. *Curr. Opin. Colloid Interface Sci.* **2010**, *15* (6), 435–444.

(33) Bysell, H.; Månsson, R.; Hansson, P.; Malmsten, M. Microgels and Microcapsules in Peptide and Protein Drug Delivery. *Adv. Drug Deliv. Rev.* **2011**, *63* (13), 1172–1185.

(34) Hansson, P.; Bysell, H.; Månsson, R.; Malmsten, M. Peptide-Microgel Interactions in the Strong Coupling Regime. *J. Phys. Chem. B* **2012**, *116* (35), 10964–10975.

(35) Melander, W. R.; El Rassi, Z.; Horváth, C. Interplay of Hydrophobic and Electrostatic Interactions in Biopolymer Chromatography. *J. Chromatogr. A* **1989**, *469*, 3–27.

(36) Roush, D. J.; Gill, D. S.; Willson, R. C. Anion-Exchange Chromatographic Behavior of Recombinant Rat Cytochrome b5. *J. Chromatogr. A* **1993**, *653* (2), 207–218.

(37) Ståhlberg, J. Retention Models for Ions in Chromatography. *J. Chromatogr. A* **1999**, *855* (1), 3–55.

(38) Hallgren, E.; Kálmán, F.; Farnan, D.; Horváth, C.; Ståhlberg, J. Protein Retention in Ion-Exchange Chromatography: Effect of Net Charge and Charge Distribution. *J. Chromatogr. A* **2000**, *877* (1–2), 13–24.

(39) Vinogradov, S. V.; Batrakova, E. V.; Kabanov, A. V. Nanogels for Oligonucleotide Delivery to the Brain. *Bioconjug. Chem.* **2004**, *15* (1), 50–60.

(40) Ballauff, M.; Likos, C. N. Dendrimers in Solution: Insight from Theory and Simulation. *Angew. Chemie Int. Ed.* **2004**, *43* (23), 2998–3020.

(41) Xu, X.; Ran, Q.; Haag, R.; Ballauff, M.; Dzubiella, J. Charged Dendrimers Revisited: Effective Charge and Surface Potential of Dendritic Polyglycerol Sulfate. *Macromolecules* **2017**, *50* (12), 4759–4769.

(42) van der Poll, D. G.; Kieler-Ferguson, H. M.; Floyd, W. C.; Guillaudeu, S. J.; Jerger, K.; Szoka, F. C.; Fréchet, J. M. Design, Synthesis, and Biological Evaluation of a Robust, Biodegradable Dendrimer. *Bioconjug. Chem.* **2010**, *21* (4), 764–773.

(43) Khandare, J.; Calderón, M.; Dagia, N. M.; Haag, R. Multifunctional Dendritic Polymers in Nanomedicine: Opportunities and Challenges. *Chem. Soc. Rev.* **2012**, *41* (7), 2824–2848.

(44) Arvizo, R. R.; Miranda, O. R.; Moyano, D. F.; Walden, C. a.; Giri, K.; Bhattacharya, R.; Robertson, J. D.; Rotello, V. M.; Reid, J. M.; Mukherjee, P. Modulating Pharmacokinetics, Tumor Uptake and Biodistribution by Engineered Nanoparticles. *PLoS One* **2011**, *6* (9), 3–8.

(45) Ruhe, J.; Ballauff, M.; Biesalski, M.; Dziezok, P.; Grohn, F.; Johannsmann, D.; Houbenov, N.; Hugenberg, N.; Konradi, R.; Minko, S.; et al. Polyelectrolytes with Defined Molecular Architecture I. *POLYELECTROLYTES WITH Defin. Mol. Archit. I* **2004**, *165*, 189–198.

(46) Ballauff, M.; Borisov, O. Polyelectrolyte Brushes. *Curr. Opin. Colloid Interface Sci.* **2006**, *11* (2006), 316–323.

(47) Wittemann, A.; Haupt, B.; Ballauff, M. Adsorption of Proteins on Spherical Polyelectrolyte Brushes in Aqueous Solution. *Physical Chemistry Chemical Physics*. 2003, pp 1671–1677.

(48) Wittemann, A.; Ballauff, M. Interaction of Proteins with Linear Polyelectrolytes and Spherical Polyelectrolyte Brushes in Aqueous Solution. *Physical Chemistry Chemical Physics*. 2006, pp 5269–5275.

(49) Chen, K.; Xu, Y.; Rana, S.; Miranda, O. R.; Dubin, P. L.; Rotello, V. M.; Sun, L.; Guo, X. Electrostatic Selectivity in Protein-Nanoparticle Interactions. *Biomacromolecules* **2011**, *12* (7), 2552–2561.

(50) Qin, L.; Xu, Y.; Han, H.; Liu, M.; Chen, K.; Wang, S.; Wang, J.; Xu, J.; Li, L.; Guo, X. β-Lactoglobulin (BLG) Binding to Highly Charged Cationic Polymer-Grafted Magnetic Nanoparticles: Effect of Ionic Strength. *J. Colloid Interface Sci.* **2015**, *460*, 221–229.

(51) Wei, Q.; Becherer, T.; Angioletti-Uberti, S.; Dzubiella, J.; Wischke, C.; Neffe, A. T.; Lendlein, A.; Ballauff, M.; Haag, R. Protein Interactions with Polymer Coatings and Biomaterials. *Angew. Chem. Int. Ed. Engl.* **2014**, *53* (31), 8004–8031.

(52) Lynch, I.; Cedervall, T.; Lundqvist, M.; Cabaleiro-Lago, C.; Linse, S.; Dawson, K. a. The Nanoparticle-Protein Complex as a Biological Entity; a Complex Fluids and Surface Science Challenge for the 21st Century. *Adv. Colloid Interface Sci.* **2007**, *134–135*, 167–174.

(53) Cedervall, T.; Lynch, I.; Lindman, S.; Berggård, T.; Thulin, E.; Nilsson, H.; Dawson, K. a; Linse, S. Understanding the Nanoparticle-Protein Corona Using Methods to Quantify Exchange Rates and Affinities of Proteins for Nanoparticles. *Proc. Natl. Acad. Sci. U. S. A.* **2007**, *104* (7), 2050–2055.

(54) Lundqvist, M.; Stigler, J.; Elia, G.; Lynch, I.; Cedervall, T.; Dawson, K. a. Nanoparticle Size and Surface Properties Determine the Protein Corona with Possible Implications for Biological Impacts. *Proc. Natl. Acad. Sci. U. S. A.* **2008**, *105* (38), 14265–14270.

(55) Oslakovic, C.; Cedervall, T.; Linse, S.; Dahlbäck, B. Polystyrene Nanoparticles Affecting Blood Coagulation. *Nanomedicine Nanotechnology, Biol. Med.* **2012**, *8* (6), 981–986.

(56) Xia, X.-R.; Monteiro-Riviere, N. a; Riviere, J. E. An Index for Characterization of Nanomaterials in Biological Systems. *Nat. Nanotechnol.* **2010**, *5* (August), 671–675.

(57) Lynch, I.; Salvati, A.; Dawson, K. a. Protein-Nanoparticle Interactions: What Does the Cell See? *Nat. Nanotechnol.* **2009**, *4* (9), 546–547.

(58) Walczyk, D.; Bombelli, F. B.; Monopoli, M. P.; Lynch, I.; Dawson, K. a. What the Cell "sees" in Bionanoscience. *J. Am. Chem. Soc.* **2010**, *132* (19), 5761–5768.

(59) Landsiedel, R.; Ma-Hock, L.; Kroll, A.; Hahn, D.; Schnekenburger, J.; Wiench, K.; Wohlleben, W. Testing Metal-Oxide Nanomaterials for Human Safety. *Adv. Mater.* **2010**, *22*, 2601–2627.

(60) Sapsford, K. E.; Algar, W. R.; Berti, L.; Gemmill, K. B.; Casey, B. J.; Oh, E.; Stewart, M. H.; Medintz, I. L. Functionalizing Nanoparticles with Biological Molecules: Developing Chemistries That Facilitate Nanotechnology. *Chem. Rev.* **2013**, *113*, 1904–2074.

(61) Treuel, L.; Nienhaus, G. U. Toward a Molecular Understanding of Nanoparticle-Protein Interactions. *Biophys. Rev.* **2012**, *4*, 137–147.

(62) Chen, C.; Li, Y.-F.; Qu, Y.; Chai, Z.; Zhao, Y. Advanced Nuclear Analytical and Related Techniques for the Growing Challenges in Nanotoxicology. *Chem. Soc. Rev.* **2013**, *42* (21), 8266–8303.

(63) Gianneli, M.; Polo, E.; Lopez, H.; Castagnola, V.; Aastrup, T.; Dawson, K. A. Label-Free in-Flow Detection of Receptor Recognition Motifs on the Biomolecular Corona of Nanoparticles. *Nanoscale* **2018**, *10* (12), 5474–5481.

(64) Lara, S.; Alnasser, F.; Polo, E.; Garry, D.; Lo Giudice, M. C.; Hristov, D. R.; Rocks, L.; Salvati, A.; Yan, Y.; Dawson, K. A. Identification of Receptor Binding to the Biomolecular Corona of Nanoparticles. *ACS Nano* **2017**, *11* (2), 1884–1893.

(65) Boselli, L.; Polo, E.; Castagnola, V.; Dawson, K. A. Regimes of Biomolecular Ultrasmall Nanoparticle Interactions. *Angew. Chemie - Int. Ed.* **2017**, *56* (15), 4215–4218.

(66) Brettschneider, F.; Tölle, M.; Von der Giet, M.; Passlick-Deetjen, J.; Steppan, S.; Peter, M.; Jankowski, V.; Krause, A.; Kühne, S.; Zidek, W.; et al. Removal of Protein-Bound, Hydrophobic Uremic Toxins by a Combined Fractionated Plasma Separation and Adsorption Technique. *Artif. Organs* **2013**, *37* (13), 409–416.

(67) Devine, E.; Krieter, D. H.; Rüth, M.; Jankovski, J.; Lemke, H. D. Binding Affinity and Capacity for the Uremic Toxin Indoxyl Sulfate. *Toxins (Basel).* **2014**, *6*, 416–430.

(68) Böhringer, F.; Jankowski, V.; Gajjala, P. R.; Zidek, W.; Jankowski, J. Release of Uremic Retention Solutes from Protein Binding by Hypertonic Predilution Hemodiafiltration. *ASAIO J.* **2015**, *61*, 55–60.

(69) Yu, S.; Schuchardt, M.; Tölle, M.; Van Der Giet, M.; Zidek, W.; Dzubiella, J.; Ballauff, M. Interaction of Human Serum Albumin with Uremic Toxins: A Thermodynamic Study. *RSC Adv.* **2017**, *7* (45), 27913–27922.

(70) Jelesarov, I.; Bosshard, H. R. Isothermal Titration Calorimetry and Differential Scanning Calorimetry as Complementary Tools to Invesitigate the Energetics of Biomolecular Recognition. *J. Mol. Recognit.* **1999**, *12*, 3–18.

(71) Chaires, J. B. Calorimetry and Thermodynamics in Drug Design. *Annu. Rev. Biophys.* **2008**, *37*, 135–151.

(72) Velázquez Campoy, A.; Freire, E. ITC in the Post-Genomic Era...? Priceless. *Biophys. Chem.* **2005**, *115* (2–3), 115–124.

(73) Klebe, G. Applying Thermodynamic Profiling in Lead Finding and Optimization. *Nat. Rev. Drug Discov.* **2015**, *14* (2), 95–110.

(74) Vega, S.; Abian, O.; Velazquez-Campoy, A. A Unified Framework Based on the Binding Polynomial for Characterizing Biological Systems by Isothermal Titration Calorimetry. *Methods* **2015**, *76*, 99–115.

(75) Ball, V.; Maechling, C. Isothermal Microcalorimetry to Investigate Non Specific Interactions in Biophysical Chemistry. *Int. J. Mol. Sci.* **2009**, *10* (8), 3283–3315.

(76) Geschwindner, S.; Ulander, J.; Johansson, P. Ligand Binding Thermodynamics in Drug Discovery: Still a Hot Tip? *Journal of Medicinal Chemistry*. 2015, pp 6321–6335.

(77) Kayitmazer, A. B. Thermodynamics of Complex Coacervation. *Adv. Colloid Interface Sci.* **2017**, *239*, 169–177.

(78) LiCata, V. J.; Liu, C. C. Analysis of Free Energy versus Temperature Curves in Protein Folding and Macromolecular Interactions. *Methods Enzymol.* **2011**, *488* (C), 219–238.

(79) Liu, C.-C.; Richard, A. J.; Datta, K.; LiCata, V. J. Prevalence of Temperature-Dependent Heat Capacity Changes in Protein-DNA Interactions. *Biophys. J.* **2008**, *94* (8), 3258–3265.

(80) Datta, K.; Wowor, A. J.; Richard, A. J.; LiCata, V. J. Temperature Dependence and Thermodynamics of Klenow Polymerase Binding to Primed-Template DNA. *Biophys. J.* **2006**, *90* (5), 1739–1751.

(81) Niedzwiecka, A.; Stepinski, J.; Darzynkiewicz, E.; Sonenberg, N.; Stolarski, R. Positive Heat Capacity Change upon Specific Binding of Translation Initiation Factor eIF4E to mRNA 5' Cap †. *Biochemistry* **2002**, *41* (40), 12140–12148.

(82) Datta, K. Thermodynamics of the Binding of Thermus Aquaticus DNA Polymerase to Primed-Template DNA. *Nucleic Acids Res.* **2003**, *31* (19), 5590–5597.

(83) Rosenfeldt, S.; Wittemann, A.; Ballauff, M.; Breininger, E.; Bolze, J.; Dingenouts, N. Interaction of Proteins with Spherical Polyelectrolyte Brushes in Solution as Studied by Small-Angle X-Ray Scattering. *Phys. Rev. E. Stat. Nonlin. Soft Matter Phys.* **2004**, *70* (6 Pt 1), 61403.

(84) Henzler, K.; Wittemann, A.; Breininger, E.; Ballauff, M.; Rosenfeldt, S. Adsorption of Bovine Hemoglobin onto Spherical Polyelectrolyte Brushes Monitored by Small-Angle X-Ray Scattering and Fourier Transform Infrared Spectroscopy. *Biomacromolecules* **2007**, *8* (11), 3674–3681.

(85) Henzler, K.; Rosenfeldt, S.; Wittemann, A.; Harnau, L.; Finet, S.; Narayanan, T.; Ballauff, M. Directed Motion of Proteins along Tethered Polyelectrolytes. *Phys. Rev. Lett.* **2008**, *100* (15), 158301.

(86) Henzler, K.; Haupt, B.; Rosenfeldt, S.; Harnau, L.; Narayanan, T.; Ballauff, M. Interaction Strength between Proteins and Polyelectrolyte Brushes: A Small Angle X-Ray Scattering Study. *Phys. Chem. Chem. Phys.* **2011**, *13* (39), 17599–17605.

(87) Yu, S.; Kent, B.; Jafta, C. J. C. J.; Petzold, A.; Radulescu, A.; Schuchardt, M.; Tölle, M.; van der Giet, M.; Zidek, W.; Ballauff, M. Stability of Human Serum Albumin Structure upon Toxin Uptake Explored by Small Angle Neutron Scattering. *Polym. (United Kingdom)* **2018**, *141*, 175–183.

(88) Anikin, K.; Röcker, C.; Wittemann, A.; Wiedenmann, J.; Ballauff, M.; Nienhaus, G. U. Polyelectrolyte-Mediated Protein Adsorption: Fluorescent Protein Binding to Individual Polyelectrolyte Nanospheres. *J. Phys. Chem. B* **2005**, *109* (12), 5418–5420.

(89) Röcker, C.; Pötzl, M.; Zhang, F.; Parak, W. J.; Nienhaus, G. U. A Quantitative Fluorescence Study of Protein Monolayer Formation on Colloidal Nanoparticles. *Nat. Nanotechnol.* **2009**, *4* (September), 577–580.

(90) Becker, A. L.; Henzler, K.; Welsch, N.; Ballauff, M.; Borisov, O. Proteins and Polyelectrolytes: A Charged Relationship. *Curr. Opin. Colloid Interface Sci.* **2012**, *17* (2), 90–96.

(91) Yu, S.; Xu, X.; Yigit, C.; van der Giet, M.; Zidek, W.; Jankowski, J.; Dzubiella, J.; Ballauff, M. Interaction of Human Serum Albumin with Short Polyelectrolytes: A Study by Calorimetry and Computer Simulations. *Soft Matter* **2015**, *11* (23), 4630–4639.

(92) Angioletti-Uberti, S.; Ballauff, M.; Dzubiella, J. Dynamic Density Functional Theory of Protein Adsorption on Polymer-Coated Nanoparticles. *Soft Matter* **2014**, *10* (40), 7932–7945.

(93) Yigit, C.; Kanduč, M.; Ballauff, M.; Dzubiella, J. Interaction of Charged Patchy Protein Models with Like-Charged Polyelectrolyte Brushes. **2017**, *33* (1), 417–427.

(94) Xu, X.; Ran, Q.; Dey, P.; Nikam, R.; Haag, R.; Ballauff, M.; Dzubiella, J. Counterion-Release Entropy Governs the Inhibition of Serum Proteins by Polyelectrolyte Drugs. *Biomacromolecules* **2018**, *19* (2), 409–416.

(95) Ran, Qidi, Xu, Xiao, Dzubiella, Pradip Dey, Shun Yu, Yan Lu, Joachim, Haag, Rainer, Ballauff, M. Interaction of Human Serum Albumin with Dendritic Polyglycerol Sulfate: Rationalizing the Thermodynamics of Binding. *J. Chem. Phys.* **2018**.

(96) Welsch, N.; Wittemann, A.; Ballauff, M. Enhanced Activity of Enzymes Immobilized in Thermoresponsive Core-Shell Microgels. *J. Phys. Chem. B* **2009**, *113* (49), 16039–16045.

(97) Haupt, B.; Neumann, T.; Wittemann, A.; Ballauff, M. Activity of Enzymes Immobilized in Colloidal Spherical Polyelectrolyte Brushes. *Biomacromolecules* **2005**, *6* (2), 948–955.

(98) Henzler, K.; Haupt, B.; Lauterbach, K.; Wittemann, A.; Borisov, O.; Ballauff, M. Adsorption of β-Actoglobulin on Spherical Polyelectrolyte Brushes: Direct Proof of Counterion Release by Isothermal Titration Calorimetry. *J. Am. Chem. Soc.* **2010**, *132* (9), 3159–3163.

(99) Leermakers, F. a M.; Ballauff, M.; Borisov, O. V. Counterion Localization in Solutions of Starlike Polyelectrolytes and Colloidal Polyelectrolyte Brushes: A Self-Consistent Field Theory. *Langmuir* **2008**, *24* (7), 10026–10034.

(100) Adroher-Benítez, I.; Moncho-Jordá, A.; Dzubiella, J. Sorption and Spatial Distribution of Protein Globules in Charged Hydrogel Particles. *Langmuir* **2017**, *33* (18), 4567–4577.

(101) Yigit, C.; Heyda, J.; Ballauff, M.; Dzubiella, J. Like-Charged Protein-Polyelectrolyte Complexation Driven by Charge Patches. *J. Chem. Phys.* **2015**, *143* (6), 64905.

(102) Record, M. T.; Anderson, C. F.; Lohman, T. M. Thermodynamic Analysis of Ion Effects on the Binding and Conformational Equilibria of Proteins and Nucleic Acids: The Roles of Ion Association or Release, Screening, and Ion Effects on Water Activity. *Q. Rev. Biophys.* **1978**, *11*, 103–178.

(103) Ball, V.; Winterhalter, M.; Schwinte, P.; Lavalle, P.; Voegel, J. C.; Schaaf, P. Complexation Mechanism of Bovine Serum Albumin and Poly(allylamine Hydrochloride). *J. Phys. Chem. B* **2002**, *106*, 2357–2364.

(104) Borisov, O. V. O. V.; Zhulina, E. B. E. B.; Leermakers, F. A. M. F. A. M.; Ballauff, M.; Müller, A. H. E. A. H. E. Conformations and Solution Properties of Star-Branched Polyelectrolytes. *Adv. Polym. Sci.* **2011**, *241* (1), 1–55.

(105) Angioletti-Uberti, S.; Ballauff, M.; Dzubiella, J. Competitive Adsorption of Multiple Proteins to Nanoparticles: The Vroman Effect Revisited. *Mol. Phys.* **2018**.

(106) Manning, G. S.; Ray, J. Counterion Condensation Revisited. *J. Biomol. Struct. Dyn.* **1998**, *16* (2), 461–476.

(107) Zhang, H.; Dubin, P. L.; Ray, J.; Manning, G. S.; Moorefield, C. N.; Newkome, G. R. Interaction of a Polycation with Small Oppositely Charged Dendrimers. *J. Phys. Chem. B* **1999**, *103* (13), 2347–2354.

(108) Deserno, M.; Holm, C.; May, S. Fraction of Condensed Counterions around a Charged Rod: Comparison of Poisson−Boltzmann Theory and Computer Simulations. *Macromolecules* **2000**, *33* (1), 199–206.

(109) Naji, A.; Jungblut, S.; Moreira, A. G.; Netz, R. R. Electrostatic Interactions in Strongly Coupled Soft Matter. *Phys. A Stat. Mech. its Appl.* **2005**, *352* (1), 131–170.

(110) Heyda, J.; Dzubiella, J. Ion-Specific Counterion Condensation on Charged Peptides: Poisson–Boltzmann vs. Atomistic Simulations. *Soft Matter* **2012**, *8* (36), 9338.

(111) Yigit, C.; Heyda, J.; Dzubiella, J. Charged Patchy Particle Models in Explicit Salt: Ion Distributions, Electrostatic Potentials, and Effective Interactions. *J. Chem. Phys.* **2015**, *143* (6), 64904.

(112) Clementi, C.; Nymeyer, H.; Onuchic, J. N. Topological and Energetic Factors: What Determines the Structural Details of the Transition State Ensemble and "en-Route" Intermediates for Protein Folding? An Investigation for Small Globular Proteins. *J. Mol. Biol.* **2000**, *298* (5), 937–953.

(113) Noel, J. K.; Whitford, P. C.; Sanbonmatsu, K. Y.; Onuchic, J. N. SMOG@ctbp: Simplified Deployment of Structure-Based Models in GROMACS. *Nucleic Acids Res.* **2010**, *38* (suppl_2), W657–W661.

(114) Hess, B.; Kutzner, C.; van der Spoel, D.; Lindahl, E. GROMACS 4: Algorithms for Highly Efficient, Load-Balanced, and Scalable Molecular Simulation. *J. Chem. Theory Comput.* **2008**, *4* (3), 435–447.

(115) General, I. J. A Note on the Standard State's Binding Free Energy. *J. Chem. Theory Comput.* **2010**, *6* (8), 2520–2524.

(116) Froehlich, E.; Mandeville, J. S.; Jennings, C. J.; Sedaghat-Herati, R.; Tajmir-Riahi, H. A. Dendrimers Bind Human Serum Albumin. *J. Phys. Chem. B* **2009**, *113* (19), 6986–6993.

(117) Ciolkowski, M.; Pałecz, B.; Appelhans, D.; Voit, B.; Klajnert, B.; Bryszewska, M. The Influence of Maltose Modified Poly(propylene Imine) Dendrimers on Hen Egg White Lysozyme Structure and Thermal Stability. *Colloids Surfaces B Biointerfaces* **2012**, *95*, 103–108.

(118) Türk, H.; Haag, R.; Alban, S. Dendritic Polyglycerol Sulfates as New Heparin Analogues and Potent Inhibitors of the Complement System. *Bioconjug. Chem.* **2004**, *15* (1), 162–167.

(119) Dernedde, J.; Rausch, A.; Weinhart, M.; Enders, S.; Tauber, R.; Licha, K.; Schirner, M.; Zugel, U.; von Bonin, A.; Haag, R. Dendritic Polyglycerol Sulfates as Multivalent Inhibitors of Inflammation. *Proc. Natl. Acad. Sci.* **2010**, *107* (46), 19679–19684.

(120) Reimann, S.; Gröger, D.; Kühne, C.; Riese, S. B.; Dernedde, J.; Haag, R. Shell Cleavable Dendritic Polyglycerol Sulfates Show High Anti-Inflammatory Properties by Inhibiting L-Selectin Binding and Complement Activation. *Adv. Healthc. Mater.* **2015**, *4* (14), 2154–2162.

(121) Boreham, A.; Pikkemaat, J.; Volz, P.; Brodwolf, R.; Kuehne, C.; Licha, K.; Haag, R.; Dernedde, J.; Alexiev, U. Detecting and Quantifying Biomolecular Interactions of a Dendritic Polyglycerol Sulfate Nanoparticle Using Fluorescence Lifetime Measurements. *Molecules* **2016**, *21* (1), 22.

(122) Woelke, A. L.; Kuehne, C.; Meyer, T.; Galstyan, G.; Dernedde, J.; Knapp, E.-W. Understanding Selectin Counter-Receptor Binding from Electrostatic Energy Computations and Experimental Binding Studies. *J. Phys. Chem. B* **2013**, *117* (51), 16443–16454.

(123) Chodera, J. D.; Mobley, D. L. Entropy-Enthalpy Compensation: Role and Ramifications in Biomolecular Ligand Recognition and Design. *Annu. Rev. Biophys.* **2013**, *42* (1), 121–142.

(124) Tomalia, D. A.; Naylor, A. M.; Goddard, W. A. Starburst Dendrimers: Molecular-Level Control

of Size, Shape, Surface Chemistry, Topology, and Flexibility from Atoms to Macroscopic Matter. *Angew. Chemie Int. Ed. English* **1990**, *29* (2), 138–175.

(125) Lin, S.-T.; Maiti, P. K.; Goddard, W. A. Dynamics and Thermodynamics of Water in PAMAM Dendrimers at Subnanosecond Time Scales. *J. Phys. Chem. B* **2005**, *109* (18), 8663–8672.

(126) Maiti, P. K.; Li, Y.; Cagin, T.; Goddard, W. A. Structure of Polyamidoamide Dendrimers up to Limiting Generations: A Mesoscale Description. *J. Chem. Phys.* **2009**, *130* (14), 144902.

(127) Welch, P.; Muthukumar, M. Tuning the Density Profile of Dendritic Polyelectrolytes. *Macromolecules* **1998**, *31* (17), 5892–5897.

(128) Carnahan, N. F.; Starling, K. E. Equation of State for Nonattracting Rigid Spheres. *J. Chem. Phys.* **1969**, *51* (2), 635–636.

(129) Lu, Y.; Ballauff, M. Thermosensitive Core-Shell Microgels: From Colloidal Model Systems to Nanoreactors. *Prog. Polym. Sci.* **2011**, *36* (6), 767–792.

(130) Wittemann, A.; Ballauff, M. Temperature-Induced Unfolding of Ribonuclease A Embedded in Spherical Polyelectrolyte Brushes. *Macromol. Biosci.* **2005**, *5* (1), 13–20.

(131) Ballauff, M.; Borisov, O. Polyelectrolyte Brushes. *Curr. Opin. Colloid Interface Sci.* **2006**, *11* (6).

(132) Wittemann, A.; Haupt, B.; Ballauff, M. Controlled Release of Proteins Bound to Spherical Polyelectrolyte Brushes. In *Zeitschrift fur Physikalische Chemie*; 2007; Vol. 221, pp 113–126.

(133) Lu, Y.; Wittemann, A.; Ballauff, M. Supramolecular Structures Generated by Spherical Polyelectrolyte Brushes and Their Application in Catalysis. *Macromol. Rapid Commun.* **2009**, *30*, 806–815.

(134) De Vos, W. M.; Leermakers, F. a M.; De Keizer, A.; Stuart, M. a C.; Kleijn, J. M. Field Theoretical Analysis of Driving Forces for the Uptake of Proteins by like-Charged Polyelectrolyte Brushes: Effects of Charge Regulation and Patchiness. *Langmuir* **2010**, *26* (14), 249–259.

(135) Wittemann, A.; Haupt, B.; Ballauff, M. Polyelectrolyte-Mediated Protein Adsorption. *Progress in Colloid and Polymer Science*. 2006, pp 58–64.

(136) Becker, A. L.; Welsch, N.; Schneider, C.; Ballauff, M. Adsorption of RNase A on Cationic Polyelectrolyte Brushes: A Study by Isothermal Titration Calorimetry. *Biomacromolecules* **2011**, *12* (11), 3936–3944.

(137) Czeslik, C.; Jackler, G.; Hazlett, T.; Gratton, E.; Steitz, R.; Wittemann, A.; Ballauff, M. Salt-Induced Protein Resistance of Polyelectrolyte Brushes Studied Using Fluorescence Correlation Spectroscopy and Neutron Reflectometry. In *Physical Chemistry Chemical Physics*; 2004; Vol. 6, pp 5557–5563.

(138) Wittemann, A.; Ballauff, M. Secondary Structure Analysis of Proteins Embedded in Spherical Polyelectrolyte Brushes by FT-IR Spectroscopy. *Anal. Chem.* **2004**, *76* (10), 2813–2819.

(139) Neumann, T.; Haupt, B.; Ballauff, M. High Activity of Enzymes Immobilized in Colloidal Nanoreactors. *Macromol. Biosci.* **2004**, *4* (1), 13–16.

(140) Henzler, K.; Haupt, B.; Ballauff, M. Enzymatic Activity of Immobilized Enzyme Determined by Isothermal Titration Calorimetry. *Anal. Biochem.* **2008**, *378* (2), 184–189.

(141) Evans, R. The Nature of the Liquid-Vapour Interface and Other Topics in the Statistical Mechanics

of Non-Uniform, Classical Fluids. *Adv. Phys.* **1979**, *28* (2), 143–200.

(142) Marconi, U. M. B.; Tarazona, P. Dynamic Density Functional Theory of Fluids. *J. Chem. Phys.* **1999**, *110* (16), 8032–8044.

(143) Vroman, L.; Adams, a. L. Findings with the Recording Ellipsometer Suggesting Rapid Exchange of Specific Plasma Proteins at Liquid/solid Interfaces. *Surf. Sci.* **1969**, *16*, 438–446.

(144) Jung, S. Y.; Lim, S. M.; Albertorio, F.; Kim, G.; Gurau, M. C.; Yang, R. D.; Holden, M. a.; Cremer, P. S. The Vroman Effect: A Molecular Level Description of Fibrinogen Displacement. *J. Am. Chem. Soc.* **2003**, *125* (14), 12782–12786.

(145) Fang, F.; Szleifer, I. Competitive Adsorption in Model Charged Protein Mixtures: Equilibrium Isotherms and Kinetics Behavior. *J. Chem. Phys.* **2003**, *119*, 1053–1065.

(146) Gong, P.; Szleifer, I. Competitive Adsorption of Model Charged Proteins: The Effect of Total Charge and Charge Distribution. *J. Colloid Interface Sci.* **2004**, *278*, 81–90.

(147) Vogler, E. a. Protein Adsorption in Three Dimensions. *Biomaterials* **2012**, *33* (5), 1201–1237.

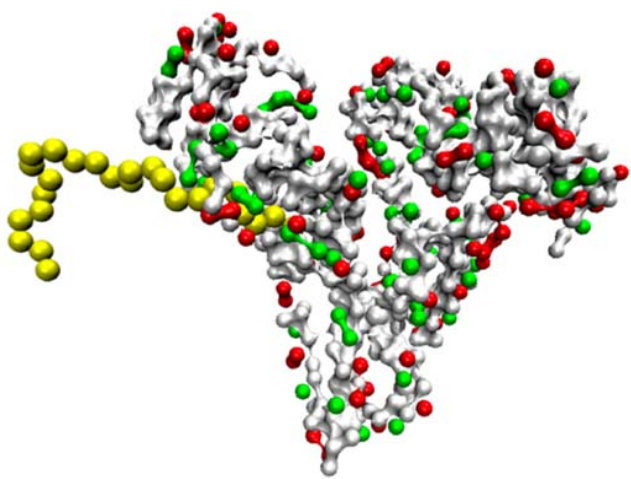

TOC graphic